\RequirePackage{rotating}
\documentclass[useAMS,usenatbib]{mn2e}

\usepackage{booktabs}
\usepackage{graphicx}
\usepackage{pdfpages}
\usepackage{txfonts}
\usepackage{natbib}
\usepackage{longtable,lscape}
\usepackage{color}
\usepackage{rotating,caption}

\newcommand\aj{AJ}
\newcommand\apj{ApJ}
\newcommand\apjs{ApJS}

\newcommand\aap{A\&A}
\newcommand\mnras{MNRAS}

\newcommand\pasa{PASA}
\newcommand\pasj{PASJ}

\newcommand\nat{Nature}

\newcommand\araa{ARA\&A}

\def\teff{\mbox{T$_{\rm eff}$}}
\def\logg{\mbox{log~{\it g}}}
\def\vmicro{\mbox{$\xi_{\rm t}$}}
\def\kmsec{\mbox{km~s$^{\rm -1}$}}

\title[Keck HIRES Spectroscopy of SkyMapper Commissioning Survey EMP
Candidates]{Keck HIRES Spectroscopy of SkyMapper Commissioning Survey Candidate Extremely Metal-Poor Stars}

\author[Marino et al.]{A. F. Marino,$^{1,2,3}$ G. S. Da Costa,$^1$ A. R. Casey,$^{1,4,5}$ M. Asplund,$^1$
M. S. Bessell,$^1$  
\newauthor
A. Frebel,$^6$ S. C. Keller,$^{1}$ K. Lind,$^{7,8}$ A. D. Mackey,$^1$ S. J. Murphy,$^{1,9}$ 
\newauthor
T. Nordlander,$^{1,10}$ J. E. Norris,$^1$ B. P. Schmidt,$^1$ and D. Yong$^1$\\ 
\\
$^{1}$Research School of Astronomy \& Astrophysics, Australian National University, ACT 2611, Australia\\
$^{2}$Dipartimento di Fisica e Astronomia ``Galileo Galilei'' - Univ. di Padova, Vicolo dell’Osservatorio 3, Padova, IT-35122\\
$^{3}$Centro di Ateneo di Studi e Attivita' Spaziali “Giuseppe Colombo” - CISAS, Via Venezia 15, Padova, IT-35131\\
$^{4}$School of Physics and Astronomy, Monash University, Wellington Rd, Clayton, VIC 3800, Australia\\
$^{5}$Faculty of Information Technology, Monash University, Wellington Rd, Clayton, VIC 3800, Australia\\
$^{6}$Department of Physics and Kavli Institute for Astrophysics and Space Research, Massachusetts Institute of Technology, Cambridge, MA 02139, USA\\
$^{7}$Max-Planck-Institut f\"{u}r Astronomie, K\"{o}nigstuhl 17, D-69117 Heidelberg, Germany\\
$^{8}$Observational Astrophysics, Department of Physics and Astronomy, Uppsala University, Box 516, SE-751 20 Uppsala, Sweden\\
$^{9}$School of Physical, Environmental and Mathematical Sciences, University of New South Wales, Canberra, ACT 2600, Australia\\
$^{10}$Center of Excellence for Astrophysics in Three Dimensions (ASTRO-3D), Australia
}

\begin{document}

\date{Accepted 2019 February 25. Received 2019 February 4; in original form 2018 December 2}

\pagerange{\pageref{firstpage}--\pageref{lastpage}} \pubyear{2017}

\maketitle

\label{firstpage}

\begin{abstract}   
We present results from the analysis of high-resolution spectra
obtained with the Keck HIRES spectrograph for a sample of 17 candidate
extremely metal-poor (EMP) stars originally selected from
commissioning data obtained with the SkyMapper telescope.  Fourteen of
the stars have not been observed previously at high dispersion.   
Three have [Fe/H] $\leq -$3.0 while the remainder, with two more
metal-rich exceptions, have $-$3.0$\leq$[Fe/H]$\leq -$2.0~dex.    
Apart from Fe, we also derive abundances for the elements C, N, Na,
Mg, Al, Si, Ca, Sc, Ti, Cr, Mn, Co, Ni, and Zn, and for
$n$-capture elements Sr, Ba, and Eu. None of the current sample of
stars is found to be carbon-rich.   
In general our chemical abundances follow previous trends found in the
literature, although we note that two of the most metal-poor stars
show very low [Ba/Fe] ($\sim -$1.7) coupled with low [Sr/Ba] ($\sim -$0.3).  
Such stars are relatively rare in the Galactic halo.  
One further star, and possibly two others, meet the criteria for
classification as a $r$-I star. 
This study, together with that of Jacobson et al.\ (2015), completes
the outcomes of the SkyMapper commissioning data survey for EMP stars.  
\end{abstract}


\section{Introduction}\label{sec:intro}

As reviewed by \citet{FN15}, the detailed study of the most metal-poor
stars in the Galaxy can provide vital clues to the processes of star
formation and to the synthesis of the chemical elements at the
earliest times.  Such stars, however, are extremely rare and at the
present time only a handful are known with [Fe/H]$\leq -$4.5~dex.  
The importance of these objects has nevertheless prompted a number of  
previous and on-going searches for such extremely metal-poor (EMP) stars 
(e.g. the HK survey \citep{BPS92}, the HES \citep{C08, AF06}, the SDSS
\citep[see][]{Aoki13}, the `Best and Brightest' survey \citep{SC14},
LAMOST \citep[see][]{Li15PASJ}, and Pristine \citep{ES17}, and references therein). 
The discovery of such stars is one of the prime science drivers behind the
SkyMapper imaging survey of the southern hemisphere sky
\citep{SK07,CW17}. The metallicity sensitivity is achieved through
the incorporation of a relatively narrow $v$-filter, whose bandpass
includes the Ca\,{\sc ii} H and K lines, into the SkyMapper filter set
\citep{MSB11}.   
The SkyMapper $uvgriz$ photometric survey of the southern sky is
ongoing \citep{CW17} but during the commissioning of the telescope a
number of $vgi$ images were taken to search for EMP candidates.
Despite the sub-optimal quality of many of the images, the program,
which we will refer to as the ``SkyMapper commissioning survey for
EMP-stars'' (to distinguish it from current on-going work) was
successful in that it resulted in the discovery of the currently
most-iron poor star known SMSS~J031300.36--670839.3 \citep{SK14,
  MSB15, TN17}.  The analysis of high dispersion spectra  
of a large sample of additional EMP-candidates selected from the
commissioning survey was presented in \citet{HJ15}.  
Here we present the final results from that survey -- the outcome of
high dispersion spectroscopic observations of a further sample of
SkyMapper EMP-candidates drawn from the commissioning survey
photometry.  
SkyMapper commissioning-era photometry was also employed in the search
for EMP stars in the Galactic Bulge \citep{Howes2015, Howes2016}.

The paper is organised as follows.  The following section describes
the target selection, the observations, and the data reduction
process.  Section~\ref{Sect3} then describes the determination of the
atmospheric parameters for the stars and the subsequent analysis to
derive the chemical abundances.  The abundance results are compared
with existing halo EMP-star studies, such as those of \citet{DY13},
\citet{VP14}, and \citet{HJ15}, in \S \ref{Sect4}. The results are
briefly summarised in \S \ref{Sect5}.

\section{Target sample and Observations}\label{sec:data}

As discussed briefly in \citet{HJ15}, the initial sample of EMP candidates was selected on the basis of
location in a 2-colour diagram in which a photometric metallicity
index $m_{i}$ = ($v-g$)$_{0} -$1.5($g-i$)$_{0}$ is plotted against ($g-i$)$_{0}$ \citep[see also][]{SK07}.  
Because of the variable quality of the commissioning-epoch data, and because of the calibration approach employed, the photometric candidate
list required additional input to identify the best candidates for
high dispersion spectroscopic follow-up. This was achieved by
obtaining low-resolution ($R$ $\approx$ 3000) 
spectra of the candidates with the WiFeS spectrograph \citep{MD10} on the ANU 2.3m telescope at Siding 
Spring Observatory.   The resulting flux calibrated spectra, which
cover the wavelength range $\sim$350--600~nm, are then compared with a
grid of MARCS 1D model atmosphere fluxes and the best-fit 
determined, as described in \citet{JEN13}. Because the spectra cover
the Paschen continuum as well as the Balmer jump and the Balmer lines
of hydrogen, the best-fit temperature and gravity are generally well  
determined.  Consistency with the temperature/gravity relation for an
old metal-poor isochrone\footnote{The isochrones employed are those
  from \citet{Vandenberg2006} supplemented with additional isochrones
  for metallicities below [Fe/H]=$—$2.3~dex (Vandenberg, priv.\ comm.\
  2009).}, which is appropriate for halo stars, provides a constraint on the adopted
reddening while the strengths of 
metal-lines such as Ca\,{\sc ii} H and K and Mg\,{\sc i} b provide the
metallicity information 
for a given temperature and gravity.  The outcome of the 2.3m
spectroscopy is then a sample of EMP candidates that can be used with
some confidence as a basis for follow-up studies at high dispersion.   

The candidates observed with the HIRES spectrograph \citep{Vogt94} at
the Keck-I telescope were those in the commissioning survey EMP
candidate sample that had not been previously observed at
high-dispersion 
\citep[cf.][]{HJ15}, that were accessible from the Keck Observatory on
the scheduled date, and that had low-resolution spectroscopic
abundance estimates \mbox{[Fe/H]$_{\mathrm {2.3m}}$ $\leq -$2.5~dex},
as determined from the 2.3m spectra.   
In all, HIRES spectra were obtained for 15 candidate EMP stars on the
ANU-allocated night of 21 September 2013 (UT), together with spectra
of two stars that had also been observed at Magellan with the MIKE
spectrograph in the \citet{HJ15} study.  One further star,
SMSS~J221334.13--072604.1, was subsequently found to be included in
the sample analysed by \citet{Aoki13} under the designation
SDSS~J2213--0726. 

Observing conditions were good with the seeing slowly rising from
0.6$\arcsec$ to 1$\arcsec$ by the  
end of the night.  The spectrograph was configured with the HIRESb
cross-disperser and the C1 decker that has a slit width of
0.86$\arcsec$ yielding a resolution $R$ $\approx$ 50,000.  Detector
binning was 2 (spatial) $\times$ 1 (spectral) and the low-gain setting
($\sim$2e$^{-}$/DN) was used for the 3 CCDs in the detector mosaic.   
Details of the observations are given in Table~\ref{tab:log}. 
The table lists the SkyMapper survey designations, the
positions, and the SkyMapper $g$, ($g-i$)$_{0}$ and $m_{i}$ photometry
taken from the SkyMapper DR1.1 data release \citep{CW17}, which
supersede the original commissioning-era photometry.   
The reddening corrections follow the procedure outlined in
\citet{CW18} while $m_{i}$ is the metallicity index, defined as 
($v-g$)$_{0} - 1.5$($g-i$)$_{0}$, for which more negative values at fixed
colour indicate potentially lower metallicity \citep[see][]{SK07, GDaC19}. 
Also given are the integration times and 
the S/N per pixel of the reduced spectra at 450~nm and 600~nm.  The median
values are 22~pix$^{-1}$ at 450~nm and 26~pix$^{-1}$ at 600~nm.

\begin{figure}
\centering
\includegraphics[width=0.34\textwidth, angle=270]{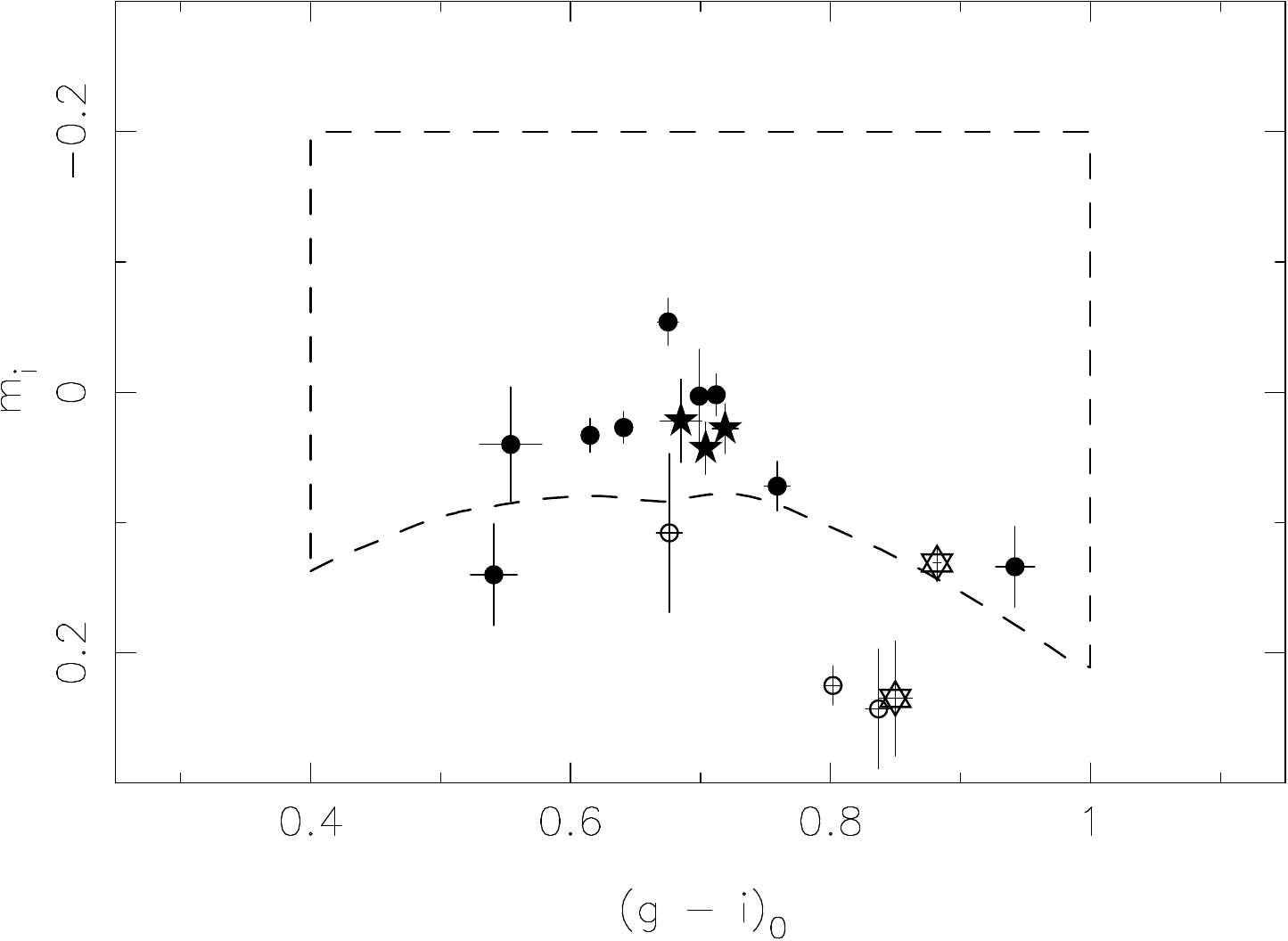} 
\caption{The location of the stars observed with Keck in the SkyMapper
  metallicity sensitive diagram using DR1.1 photometry. The selection
  window currently employed to select candidate EMP stars is shown by
  the dashed boundary.  Stars whose abundances are determined here to
  have [Fe/H]$_{\mathrm {LTE}}$$\leq -$3.0 are plotted as filled
  star-symbols, those with $-$3.0 $\leq$[Fe/H]$_{\mathrm {LTE}}$$\leq
  -$2.5 are plotted as filled circles, 
  open circles show stars with $-$2.5 $\leq$[Fe/H]$_{\mathrm
    {LTE}}$$\leq -$2.0 and the six-point star symbol shows the star
  for which [Fe/H]$_{\mathrm {LTE}}$$> -$2.0~dex. Individual error bars are
  shown. These have been calculated from the photometry errors given
  in the DR1.1 database.}  
\label{fig:mi_vs_gi}
\end{figure}

In Fig.\ \ref{fig:mi_vs_gi} we show the location of the observed stars
in the SkyMapper metallicity-sensitive diagram based on 
the DR1.1 photometry. Shown also in the figure is the selection
window that is used in defining photometric EMP candidates for the
current (post commissioning) survey, where the lower boundary is set
by the location of the [M/H]=$-$2.0~dex, 12.5~Gyr isochrone in this
plane \citep[see][in prep. for details]{GDaC19}. While photometric
uncertainties, particularly in the $v$-magnitudes, introduce scatter
in this diagram, it is reassuring that all but one of the 12
candidates which are found in the analysis here to have [Fe/H]$_{\mathrm {LTE}}$
$\leq -$2.5~dex 
(where LTE means that the Fe abundance is obtained assuming the
  local thermodynamic equilibrium)  
are within the selection window 
while there is only one contaminant -- a star found here to have
[Fe/H]$_{\mathrm {LTE}}$ $> -$2.0 despite lying (just) in the selection
region. Although the sample is small, Fig.\ \ref{fig:mi_vs_gi} does
verify that the current SkyMapper photometric selection process
efficiently finds stars with [Fe/H]$_{\mathrm {LTE}}$ $\leq -$2.5
with only a very minor degree of contamination. 
In fact, \citet{GDaC19} show that in the current
on-going program, $\sim$85\% of the SkyMapper DR1.1 photometric EMP
candidates that lie within the selection window shown in
Fig.~\ref{fig:mi_vs_gi}, and which also possess metallicity estimates 
from low resolution 2.3m spectra, have [Fe/H]$_{2.3m}$ $\leq -$2.0~dex,
while $\sim$40\% have [Fe/H]$_{2.3m}$$\leq -$2.75~dex. The best
candidates are then followed-up at high dispersion with the MIKE
echelle spectrograph on the Magellan 6.5m telescope. 

The observed spectra were processed with the standard HIRES reduction
pipeline MAKEE to obtain flat-fielded, extracted,
wavelength-calibrated, velocity-corrected spectra for 
each echelle order. For the subsequent analysis the individual
spectral orders were merged into a single continuous spectrum for each
of the 3 CCD detectors, which was then continuum normalized and
wavelength-offset by the observed geocentric velocity. 

Radial velocities (RVs) were derived using the IRAF@FXCOR task,
which cross-correlates the object spectrum with a template
spectrum. For the template, we used a synthetic spectrum obtained
through the June 2014 version of MOOG \citep{MOOG73}. This spectrum
was computed with a stellar model atmosphere interpolated from the
\citet{CK04} grid, adopting parameters (effective
temperature \teff, surface gravity \logg, microturbulence \vmicro,
metallicity [M/H]) = (4800~K, 1.5, 2.0~\kmsec, $-$2.50).  
The errors associated to RVs due to the cross-correlation technique are generally
small, in our case are between $\sim$0.2 and $\sim$0.6~\kmsec. 
As we do not have repeated observations of the same star, we cannot
provide more realistic estimates of the internal velocity
uncertainties. 

Independent radial velocity measurements are available for four of our
stars: the two in common with \citet{HJ15} (SMSS~J010839.58--285701.5
and SMSS~J034249.52--284215.8), the star in common with \citet{Aoki13},
and the star SMSS~J202059.17--043447.0 which has a radial velocity
tabulated in Gaia DR2 \citep{Gaia}.  SMSS~J010839.58--285701.5 also has
a radial velocity listed in Gaia DR2. Comparison with these
independent values reveals that a correction of $—$84$\pm$2 (standard error
of the mean) \kmsec\ to our velocities is required for agreement.  We
can find no obvious explanation for this velocity offset but have
verified its existence via an independent reduction of a subset of the
observed spectra.  With the offset applied our velocities agree well
with the published values, and there is no evidence for any
significant velocity variability in these four stars.  Table 1 then
lists the heliocentric radial velocity for each star in our sample
after applying the velocity offset.  Many of the stars have large
heliocentric velocities as expected for a sample dominated by Galactic
halo stars.

\begin{table*}
\caption{Details of the target stars and the Keck HIRES observations.}
\label{tab:log}
\scriptsize
\begin{center}
\begin{tabular}{ccccrrccccc}
\hline
ID & RA      & Dec     & $g$ & ($g-i$)$_{0}$ & $m_{i}$ & Exp Time & ~~S/N~[450~nm] & ~~S/N~[600~nm]& RV       & \\
   & (J2000) & (J2000) &     &               &        & (sec)    & (pix$^{-1}$)   & (pix$^{-1}$) & (\kmsec) & Notes \\
\hline
SMSS~J202059.17--043447.0 & 20 20 59.17 & $-$04 34 47.0 & 13.815 & 0.837 & 0.243    & ~~660 & 34 & 38 & $-$190.7 &  \\
SMSS~J202400.03--024445.9 & 20 24 00.03 & $-$02 44 45.9 & 15.631 & 0.850 & 0.235    & ~~660 & 15 & 17 & $-$133.8 &  \\
SMSS~J202601.44--033002.5 & 20 26 01.44 & $-$03 30 02.5 & 15.508 & 0.541 & 0.140    & 1800  & 21 & 26 & $-$149.5 &  \\
SMSS~J202659.41--031149.2 & 20 26 59.41 & $-$03 11 49.2 & 14.180 & 0.641 & 0.027    & ~~600 & 28 & 29 & $-$199.1 &  \\
SMSS~J204654.92--020409.2 & 20 46 54.92 & $-$02 04 09.2 & 15.026 & 0.882 & 0.131    & 1200  & 26 & 30 & $-$451.2 &  \\
SMSS~J211657.83--012517.5 & 21 16 57.83 & $-$01 25 17.5 & 14.515 & 0.675 & $-$0.054 & ~~900 & 21 & 22 & $-$155.4 &  \\
SMSS~J212001.71--001158.6 & 21 20 01.71 & $-$00 11 58.6 & 15.582 & 0.942 & 0.134    & 1500  & 22 & 26 & $-$350.2 &  \\
SMSS~J212113.63--005132.2 & 21 21 13.63 & $-$00 51 32.2 & 14.854 & 0.712 & 0.002    & 1200  & 21 & 23 &~$+$33.9  &  \\
SMSS~J212217.52--295552.7 & 21 22 17.52 & $-$29 55 52.7 & 16.387 & 0.554 & 0.040    & 2700  & 18 & 19 & $+$200.8 &  \\
SMSS~J220514.10--013407.6 & 22 05 14.10 & $-$01 34 07.6 & 13.933 & 0.802 & 0.225    & ~~600 & 22 & 26 & $-$199.2 &  \\
SMSS~J220535.05--004403.6 & 22 05 35.05 & $-$00 44 03.6 & 14.214 & 0.704 & 0.043    & ~~600 & 25 & 28 &~$-$62.2  &  \\
SMSS~J221334.13--072604.1 & 22 13 34.13 & $-$07 26 04.1 & 15.310 & 0.759 & 0.072    & 1500  & 19 & 22 & $-$392.7 &1 \\
SMSS~J222349.48--114751.1 & 22 23 49.48 & $-$11 47 51.1 & 15.177 & 0.719 & 0.028    & 1500  & 22 & 23 & $-$367.1 &  \\
SMSS~J225336.83--270435.4 & 22 53 36.83 & $-$27 04 35.4 & 16.024 & 0.685 & 0.022    & 1500  & 16 & 17 & $+$116.6 &  \\
SMSS~J230306.22--041621.8 & 23 03 06.22 & $-$04 16 21.8 & 14.606 & 0.699 & 0.003    & 1200  & 26 & 28 & $-$244.6 &  \\
SMSS~J010839.58--285701.5 & 01 08 39.58 & $-$28 57 01.5 & 12.747 & 0.615 & 0.033    & ~~300 & 34 & 35 & $+$144.3 &2 \\
SMSS~J034249.52--284215.8 & 03 42 49.52 & $-$28 42 15.8 & 14.646 & 0.676 & 0.108    & 1200  & 29 & 31 & $+$157.2 &2 \\
\hline
\end{tabular}
\end{center}
Notes:
(1) In \citet{Aoki13} as SDSS J2213--0726;
(2) In \citet{HJ15}
\end{table*}

\section{Chemical abundances analysis} \label{Sect3}

Chemical abundances were derived from a local thermodynamic
equilibrium (LTE) analysis by using the June 2014 
version of the spectral analysis code MOOG 
\citep{MOOG73}, together with the alpha-enhanced Kurucz model
atmospheres of \citet{CK04}, whose parameters have been obtained as
described in Sect.~\ref{sec:atm}.  The reference solar abundances
adopted were those of \citet{MA09}.  
 
In the following sections we detail the approach employed to derive
the adopted atmospheric parameters, and describe the spectral features
used to infer the chemical abundances.  In general, we follow the
procedures outlined in \citet{HJ15} in order to facilitate a direct
comparison of the results obtained here with those in that work. 

\subsection{Atmospheric parameters}\label{sec:atm}

The atmospheric parameters were derived via a number of different steps.
First, as in \citet{HJ15}, initial values of \teff\ and the
microturbulence \vmicro\ were determined by imposing excitation 
potential (E.P.) equilibrium for Fe\,{\sc i}, to yield \teff, and by
removing any trend between Fe\,{\sc i} abundance and the reduced
equivalent width (EW) to fix \vmicro.  For the majority of the stars
observed, however, there was a paucity of measureable Fe\,{\sc ii}
lines in the spectra invalidating the determination of a spectroscopic
\logg\ value by matching Fe\,{\sc i} and Fe\,{\sc ii} abundances.
Instead, we derived \logg\ by matching the \teff\ value with a 12~Gyr
Yonsei-Yale isochrone \citep{PD04} that has [$\alpha$/Fe]=$+$0.4 and
the appropriate metallicity for the star as derived from the initial
analysis.  The procedure was then iterated until the \teff, \logg\ and
\vmicro\ values did not change appreciably -- usually only one
iteration was required. 

However, as noted by \citet[][and the references therein]{HJ15},
spectroscopic \teff\ values for metal-poor red giants are  
generally cooler than those inferred from photometry due to departures from LTE\@.
\citet{HJ15} dealt with this issue by adopting corrections to the
spectroscopic effective temperatures as described in 
\citet{AF13}. These corrections shift the spectroscopic temperatures
to a scale that is more consistent with photometrically derived
temperatures. Such a shift is also supported by the recent detailed 3D
non-LTE calculations for Fe \citep{Amarsi2016} and H
\citep{Amarsi2018}. Consequently, in order to allow direct comparison
of the abundances derived here with those of \cite{HJ15}, we have
followed the same approach: the spectroscopic \teff\ values have been
corrected as described in \citet{AF13} leading to updated values of
\vmicro\ and the isochrone-based \logg\ values. Again the process was
iterated until convergence was achieved, and the resulting values used
in the abundance analysis.  
In the end, the corrections applied to the spectroscopic \teff\
ranged between $\sim$150 and $\sim$220~K, being larger for the
cooler stars.

\begin{figure}
\centering
\includegraphics[width=0.48\textwidth]{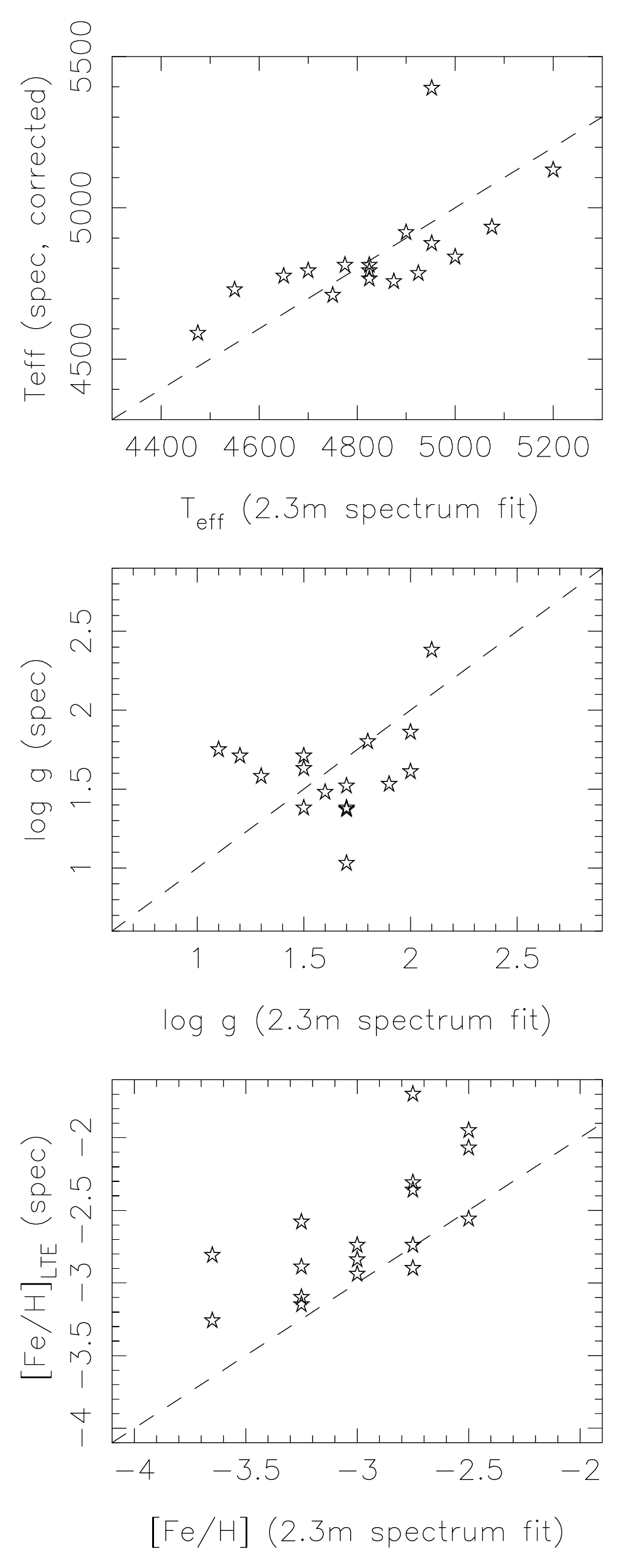}
\caption{{\it Top panel}: the adopted corrected spectroscopic temperatures are shown as a function of the temperatures obtained from spectral fits to the low-resolution 2.3m spectra.  The dashed line is the 1:1 relation. {\it Middle panel}: the adopted spectroscopic gravities are plotted against the \logg\ values derived from the fits to the low-resolution spectra.  The dashed line is the 1:1 relation.  {\it Bottom panel}: a comparison of the [Fe/H]$_{\mathrm {LTE}}$ values determined from the Keck spectra with those estimated from the low-resolution spectral fits.  The dashed line is again a 1:1 relation.}
\label{f2:t_g_comp}
\end{figure}

We can verify the suitability of the final adopted stellar parameters
by comparing them with the \teff\ and \logg\ values derived from the
spectrophotometric fits to the 2.3m low-resolution spectra.  The
comparison is shown in the upper panels 
of Fig.\ \ref{f2:t_g_comp}.  The top panel compares the corrected
spectroscopic \teff\ values with the spectrophotometric
determinations: it shows excellent agreement -- the points scatter
about the 1:1 line and the mean difference between the determinations
is only 10 K (spectroscopic \teff\ hotter) with a standard deviation
of 150 K.  Ascribing equal uncertainties to each method then indicates
that the uncertainty in the adopted spectroscopic \teff\ values is of
order 100 K\@.  
The largest discrepancy occurs for star SMSS~J212217.52--295552.7
where the corrected spectroscopic temperature is $\sim$450~K hotter
than the spectrophotometric determination. There is no
straightforward explanation for this difference although we note
that, based on the other stars in the sample, the spectrophotometric
temperature for this star is too cool by $\sim$200~K for its
($g-i$)$_{0}$ colour. We also note that with $g \approx$16.4, this
star is fainter than the usual $g$=16 limit for 2.3m follow-up
observations, while the HIRES observations have one of the lowest
S/N values in the sample.  For consistency of approach we retain the
use of the corrected spectroscopic temperature for this star,
although the uncertainty is likely larger than the typical
$\pm$100~K value. 

The middle panel shows the comparison for the \logg\ values.  
Here the mean difference, in the sense \mbox{log~{\it
    g$_{2.3m}$}}$-$\mbox{log~{\it g$_{spec}$}}, is $+$0.05~dex with a 
standard deviation of 0.35~dex after excluding  
SMSS~J212217.52--295552.7 where the large difference in temperature and
our isochrone-based approach to fix \logg$_{\mathrm spec}$, results in
a significant offset from the spectrophotometric value. 
Again assuming equal uncertainties in each method, this suggests that
the uncertainty in the adopted \logg\ values is of order of $\sim$0.25~dex.
The adopted atmospheric parameters and the resulting [Fe/H]$_{\mathrm
  {LTE}}$ values are given in Table~\ref{tab2_data}. 
Together with the [Fe/H]$_{\mathrm {LTE}}$ we list the
  [Fe/H]$_{\mathrm {non-LTE}}$ values obtained by applying non-LTE
  corrections to Fe~{\sc {i}} lines as in \citet{KL12, KL17}. For
  comparison reasons, in the following we will use our
  [Fe/H]$_{\mathrm {LTE}}$ values.

An independent check on the adopted atmospheric parameters is provided
by a comparison with those adopted in \citet{HJ15} for the two stars
in common. For SMSS~J010839.58--285701.5 we find
\teff/\logg/[Fe/H]$_{\mathrm {LTE}}$ values of 4936/1.86/$-$2.90, while
\cite{HJ15} list 4855/1.55/$-$3.02.  Similarly, for
SMSS~J034249.52--284215.8 we find 4783/1.58/$-$2.31 compared to   
4828/1.60/$-$2.33 in \citet{HJ15}. The differences in the parameters
are reassuringly low giving confidence that the results derived here
can be straightforwardly compared with those of \citet{HJ15}. We also
note that for star 
SMSS~J221334.13--072604.1, \citet{Aoki13} list parameters of
5150/1.8/$-$2.55 while we find 4810/1.52/$-$2.89; the higher abundance
given by \citet{Aoki13} is likely largely a direct consequence of the
more than 300~K higher temperature employed in that study. 
For completeness, we note that \citet{Aoki13} did not determine
spectroscopic temperatures, rather they used the temperatures
determined by the SEGUE Stellar Parameter Pipeline (SSPP) from the
SEGUE low resolution spectra - see \citet{Lee11}  and references
therein.  For this particular star, however, the temperature
estimates given (but not used) by \citet{Aoki13} from the
($V-K$)$_{0}$ and ($g-r$)$_{0}$ colours, namely 4724~K and 4867~K,
are much more consistent with our determination of 4810~K than the
SSPP value used by \citet{Aoki13}.

For completeness we also show in the bottom panel of
Fig.~\ref{f2:t_g_comp} a comparison between the [Fe/H]$_{2.3m}$
values estimated from the fits to the low-resolution spectra and the final
[Fe/H]$_{\mathrm {LTE}}$ values determined from the analysis of the
high-resolution Keck spectra.  Given that the low-resolution values
are quantized at the 0.25 dex level, the agreement is reasonable: the
mean difference is 0.33 dex, with the low-resolution estimates being
lower, and the standard deviation of the differences is 0.32 dex.

In the following estimates of the internal uncertainties in chemical
abundances due to the adopted model atmospheres will be estimated by
varying the stellar parameters, one at a time, by
\teff/\logg/[M/H]/\vmicro=$\pm$100\,K/$\pm$0.40\,cgs/$\pm$0.30\,dex/$\pm$0.40\,\kmsec.

%
%

\begin{table}
\caption{Adopted atmospheric parameters \teff/\logg/[Fe/H]$_{\mathrm
    {LTE}}$/\vmicro.  [Fe/H]$_{\mathrm {non-LTE}}$ obtained values are
  also listed.} 
\label{tab2_data}
\scriptsize
\begin{center}
\begin{tabular}{cccccc}
\hline
ID &  \teff & \logg & [Fe/H]$_{\mathrm {LTE}}$ & [Fe/H]$_{\mathrm {non-LTE}}$ & \vmicro\\
   & K      & dex   & dex                    & dex                       & \kmsec\\
\hline
SMSS~J202059.17--043447.0  & 4711 & 1.38 & $-$2.30 & $-$2.17  & 1.64\\
SMSS~J202400.03--024445.9  & 4792 & 1.71 & $-$1.97 & $-$1.88  & 1.98\\
SMSS~J202601.44--033002.5  & 5125 & 2.38 & $-$2.84 & $-$2.72  & 1.55\\
SMSS~J202659.41--031149.2  & 4918 & 1.80 & $-$2.94 & $-$2.78  & 1.74\\
SMSS~J204654.92--020409.2  & 4729 & 1.75 & $-$1.71 & $-$1.62  & 1.70\\
SMSS~J211657.83--012517.5  & 4837 & 1.61 & $-$2.56 & $-$2.42  & 1.95\\
SMSS~J212001.71--001158.6  & 4585 & 1.03 & $-$2.58 & $-$2.43  & 2.45\\
SMSS~J212113.63--005132.2  & 4810 & 1.53 & $-$2.75 & $-$2.58  & 1.95\\
SMSS~J212217.52--295552.7  & 5395 & 3.17 & $-$2.81 & $-$2.73  & 2.15\\
SMSS~J220514.10--013407.6  & 4774 & 1.63 & $-$2.07 & $-$1.97  & 1.80\\
SMSS~J220535.05--004403.6  & 4765 & 1.38 & $-$3.10 & $-$2.91  & 1.85\\
SMSS~J221334.13--072604.1  & 4810 & 1.52 & $-$2.89 & $-$2.72  & 1.97\\
SMSS~J222349.48--114751.1  & 4756 & 1.37 & $-$3.15 & $-$2.96  & 1.86\\
SMSS~J225336.83--270435.4  & 4882 & 1.71 & $-$3.26 & $-$3.07  & 2.32\\
SMSS~J230306.22--041621.8  & 4792 & 1.48 & $-$2.74 & $-$2.58  & 1.72\\
SMSS~J010839.58--285701.5  & 4936 & 1.86 & $-$2.90 & $-$2.76  & 1.68\\
SMSS~J034249.52--284215.8  & 4783 & 1.58 & $-$2.31 & $-$2.19  & 1.70\\
\hline
\end{tabular}
\end{center}
\end{table}

\subsection{Chemical species analysed}

A list of the spectral lines used in the abundance analysis, together
with the excitational potentials (E.P.), the total oscillator
strengths (log~$gf$) employed, and the measured equivalent widths
(EWs) is provided in Tab.~\ref{tab:linelist}. The atomic data come
from \citet{HJ15}, with the exception of a few lines
highlighted in Tab.~\ref{tab:linelist}.
In most cases the analysis is based on the measurement of EWs via
gaussian fits to the profiles of well-isolated lines, as described in
\citet{AFM08}; exceptions to this approach are discussed below.  When
required, and when atomic data is available from the literature,
hyperfine and/or isotopic splitting was incorporated in the analysis,
as indicated in the last column of Table~\ref{tab:linelist}.  

We now comment in detail on the transitions used in the analyses for
different element classes, noting that for some species abundances are
determined only for a subset of the sample depending on the S/N of the
spectrum and the adopted atmospheric parameters.     

\begin{table}
\caption{Atomic data and equivalent widths for program stars}\label{tab:linelist}
\begin{tabular}{c c c r c}
\hline
Wavelength [\AA] & Species & E.P. [eV] & log(gf) & EW [m\AA]\\
\hline
\multicolumn{5}{c}{SMSS~J202059.17--043447.0} \\
4447.720  & 26.0 & 2.220 &   $-$1.340     & 80.0  \\
4595.360  & 26.0 & 3.290 &   $-$1.760     & 19.2  \\
4602.940  & 26.0 & 1.490 &   $-$2.210     & 90.0  \\
4630.120  & 26.0 & 2.280 &   $-$2.590     & 20.9  \\
4733.590  & 26.0 & 1.490 &   $-$2.990     & 55.0  \\
4736.770  & 26.0 & 3.210 &   $-$0.750     & 60.9  \\
4871.320  & 26.0 & 2.870 &   $-$0.360     & 91.7  \\
4903.310  & 26.0 & 2.880 &   $-$0.930     & 69.4  \\
4918.990  & 26.0 & 2.850 &   $-$0.340     & 91.1  \\
4924.770  & 26.0 & 2.280 &   $-$2.110     & 51.4  \\
4938.810  & 26.0 & 2.880 &   $-$1.080     & 56.7  \\
4939.690  & 26.0 & 0.860 &   $-$3.250     & 77.7  \\
4966.090  & 26.0 & 3.330 &   $-$0.870     & 45.0  \\
4994.130  & 26.0 & 0.920 &   $-$2.970     & 88.6  \\
5014.940  & 26.0 & 3.940 &   $-$0.300     & 37.7  \\
5044.210  & 26.0 & 2.850 &   $-$2.020     & 13.8  \\
\hline
\end{tabular}
Notes:
Only a portion of this table is shown here to demonstrate its form and
content. A machine-readable version of the full table will be available.
\end{table}

\subsubsection{Light elements} 
Carbon abundances were derived via spectral synthesis of the CH G-band
$(A^{2}\Delta - X^{2}\Pi)$ heads near 4312 and 4323~\AA.  An
oxygen abundance ratio of [O/Fe] = $+$0.4 dex was assumed as the S/N
of the spectra does not allow  a determination of the oxygen abundance
from the forbidden [O\,{\sc i}] lines at 6300 and 6363~\AA.
Examples of the  
synthetic spectrum fits are shown in the left panels of Fig.\
\ref{spfits}. Similarly, nitrogen abundances come from spectral
synthesis of the CN bands $B^{2}\Sigma-X^{2}\Sigma$ at $\sim$3880~\AA\
and $\sim$4215~\AA, using the
carbon abundance derived from the G-band fits.   
Sodium abundances were inferred from the Na resonance doublet at
$\sim$5893~\AA. For three stars we were able to estimate Na
from the doublet $\sim$5685~\AA.
Sodium abundances were then corrected for NLTE effects, as
in \citet{KL11}, and listed in Tab.~\ref{tab:abuCNNa}.
For most stars, we were able to infer Al abundances from the
spectral synthesis of the lines used also in \citet{HJ15}, namely at
$\sim$3961~\AA\ and $\sim$3944~\AA.

\begin{table*}
\caption{Chemical abundance ratios for the light elements C, N, Na and
  Al. For carbon we also list the evolutionary corrected
  [C/Fe]$_{evol}$ abundances derived by applying corrections from V.\
  Placco (private comm.); the estimated natal [C/Fe] corresponds to
  [C/Fe]$_{evol}$. For sodium we give both LTE and NLTE abundances. We
  also report the number (\#) of spectral features analysed for each
  element as well as the standard deviations of the abundances ($\sigma$). } 
\label{tab:abuCNNa}
\begin{center}
\scriptsize
\begin{tabular}{c cccr ccc ccccc ccc}
\hline
ID &  [C/Fe] & $\sigma$ & \# & [C/Fe]$_{evol}$ & [N/Fe] & $\sigma$ &\# & [Na/Fe]$_{\mathrm {LTE}}$ & $\sigma$  & [Na/Fe]$_{\mathrm {NLTE}}$ & $\sigma$  & \# & [Al/Fe] & $\sigma$ & \# \\
\hline
SMSS~J202059.17--043447.0 &$-$0.20  & 0.01 & 2 & 0.39	 & 0.92 &  0.24	  & 2 &  \ \ 0.22  & 0.18  &$-$0.04   & 0.04 & 3 & $-$0.44  & --   & 1  \\
SMSS~J202400.03--024445.9 &$-$0.40  & 0.03 & 2 & $-$0.09 & 0.94 &  0.16	  & 2 &  \ \ 0.19  & 0.08  &$-$0.20   & 0.06 & 2 & $-$1.00  & --   & 1  \\
SMSS~J202601.44--033002.5 &\ \ 0.18 & 0.02 & 2 & 0.19	 & --   &  --	  & - &  \ \ 0.27  & 0.09  &$-$0.09   & 0.03 & 2 & $-$0.68  & 0.11 & 2  \\
SMSS~J202659.41--031149.2 &$-$0.06  & 0.03 & 2 & 0.07	 & --   &  --	  & - &  \ \ 0.44  & 0.03  & \ \ 0.04 & 0.01 & 2 & $-$0.54  & 0.23 & 2  \\
SMSS~J204654.92--020409.2 &$-$0.42  & 0.01 & 2 & $-$0.15 & 0.53 &  0.08	  & 2 &   $-$0.28  & 0.10  &$-$0.46   & 0.04 & 4 &  --      &  --  & -  \\
SMSS~J211657.83--012517.5 &\ \ 0.05 & 0.02 & 2 & 0.43	 & --   &  --	  & - &  \ \ 0.23  & 0.10  &$-$0.16   & 0.10 & 2 & $-$0.83  & 0.15 & 2  \\
SMSS~J212001.71--001158.6 &$-$0.49  & 0.02 & 2 & 0.25	 & 1.13 &  0.34	  & 2 &   $-$0.05  & 0.03  &$-$0.34   & 0.04 & 2 &  --      &  --  & -  \\
SMSS~J212113.63--005132.2 &\ \ 0.26 & 0.03 & 2 & 0.66	 & --   &  --	  & - &  \ \ 0.45  & 0.03  & \ \ 0.05 & 0.01 & 2 & $-$0.40  & 0.35 & 2  \\
SMSS~J212217.52--295552.7 &\ \ 0.64 & 0.05 & 2 & 0.64	 & --   &  --	  & - &  \ \ 0.38  & 0.08  & \ \ 0.00 & 0.09 & 2 &  --      &  --  & -  \\
SMSS~J220514.10--013407.6 &$-$0.09  & 0.03 & 2 & 0.27	 & 0.80 &  0.16   & 2 &  \ \ 0.07  & 0.18  &$-$0.21   & 0.08 & 3 & $-$0.96  &  --  & 1  \\
SMSS~J220535.05--004403.6 &$-$0.01  & 0.01 & 2 & 0.51	 & --   &  --	  & - &  \ \ 0.24  & 0.17  &$-$0.02   & 0.09 & 2 & $-$0.67  & 0.18 & 2  \\
SMSS~J221334.13--072604.1 &$-$0.58  & 0.01 & 2 & $-$0.15 & --   &  --	  & - &  \ \ 0.04  & 0.05  &$-$0.22   & 0.01 & 2 & $-$0.60  &  --  & 1  \\
SMSS~J222349.48--114751.1 &$-$0.22  & 0.01 & 2 & 0.31	 & --   &  --	  & - &  \ \ 0.73  & 0.13  & \ \ 0.36 & 0.08 & 2 & $-$0.65  & 0.14 & 2  \\
SMSS~J225336.83--270435.4 &\ \ 0.20 & 0.03 & 2 & 0.37	 & --   &  --	  & - &  \ \ 0.13  & 0.08  &$-$0.07   & 0.02 & 2 & $-$0.63  &  --  & 1  \\
SMSS~J230306.22--041621.8 &\ \ 0.12 & 0.02 & 2 & 0.56	 & --   &  --	  & - &  \ \ 0.22  & 0.04  &$-$0.12   & 0.08 & 2 & $-$0.10  &  --  & 1  \\
SMSS~J010839.58--285701.5 &\ \ 0.12 & 0.01 & 2 & 0.22	 & --   &  --	  & - &  \ \ 0.71  & 0.11  & \ \ 0.23 & 0.06 & 2 & $-$0.22  & 0.09 & 2  \\
SMSS~J034249.52--284215.8 &$-$0.41  & 0.01 & 2 & 0.03    & --   &  --	  & - &   $-$0.44  & 0.07  &$-$0.71   & 0.03 & 2 & $-$1.34  & 0.49 & 2  \\
\hline
\end{tabular}
\end{center}
\end{table*}

\subsubsection{$\alpha$-elements} 
We determined chemical abundances for the $\alpha$-elements Mg, Si, Ca,
and Ti.  For magnesium, silicon and titanium 
the abundances could be determined for all the stars in the observed
sample, since at least one up to three strong lines were available for
these elements; a larger number of lines were generally detectable for
Ti, particularly for Ti\,{\sc ii}.  Calcium abundances were inferred
from only one or two lines -- see Table \ref{tab:abuALPHA}.

%
\begin{table*}
\caption{Chemical abundances for $\alpha$ elements, Mg, Si, Ca, and
  Ti. For each element we report the number (\#) of analysed spectral
  features and the resulting standard deviations ($\sigma$).}
\label{tab:abuALPHA}
\begin{center}
\begin{tabular}{c rcc rcc rcc rcc rcc}
\hline
ID &  [Mg/Fe] & $\sigma$ & \# & [Si/Fe] & $\sigma$ & \# & [Ca/Fe] & $\sigma$ &\# & [Ti\,{\sc i}/Fe] & $\sigma$ &\# &[Ti\,{\sc ii}/Fe] & $\sigma$ &\# \\
\hline
SMSS~J202059.17--043447.0     &    0.40  & 0.05  & 3 & 0.43 & 0.18 & 2 &    0.35 &  0.19 & 2 &    0.10  & 0.07  &   6 &   0.37  &  0.10   &   8\\
SMSS~J202400.03--024445.9     &    0.62  & 0.23  & 3 & 0.42 & 0.01 & 2 &    0.57 &  0.30 & 2 &    0.18  & 0.09  &   5 &   0.35  &  0.13   &   8\\
SMSS~J202601.44--033002.5     &    0.42  & 0.01  & 2 & 0.57 & 0.19 & 2 &    0.46 &  --   & 1 &    0.45  & 0.19  &   2 &   0.10  &  0.13   &   5\\
SMSS~J202659.41--031149.2     &    0.42  & 0.12  & 2 & 0.75 & 0.32 & 2 &    0.39 &  0.01 & 2 &    0.39  & 0.16  &   2 &   0.43  &  0.17   &   7\\
SMSS~J204654.92--020409.2     &    0.18  & 0.16  & 3 & 0.18 & 0.15 & 2 &    0.33 &  0.16 & 3 & $-$0.02  & 0.06  &   7 &   0.23  &  0.12   &   8\\
SMSS~J211657.83--012517.5     &    0.47  & 0.07  & 2 & 0.56 & 0.43 & 2 &    0.34 &  0.02 & 2 &    0.30  & 0.15  &   4 &   0.03  &  0.18   &   8\\
SMSS~J212001.71--001158.6     &    0.36  & 0.12  & 2 & 0.69 & 0.15 & 2 &    0.36 &  0.07 & 2 &    0.20  & 0.08  &   5 &   0.25  &  0.15   &   8\\
SMSS~J212113.63--005132.2     &    0.35  & 0.04  & 2 & 0.64 & 0.46 & 2 &    0.33 &  0.21 & 2 &    0.38  & 0.01  &   2 &   0.00  &  0.15   &   7\\
SMSS~J212217.52--295552.7     &    0.21  & 0.09  & 2 & 0.87 & --   & 1 &    --   &  --   & - &    --    & --    &   - &$-$0.21  &  0.04   &   2\\
SMSS~J220514.10--013407.6     &    0.43  & 0.16  & 3 & 0.28 & 0.33 & 2 &    0.37 &  0.03 & 2 &    0.22  & 0.04  &   6 &   0.31  &  0.13   &   8\\
SMSS~J220535.05--004403.6     &    0.26  & 0.19  & 2 & 0.72 & 0.36 & 2 &    0.21 &  --   & 1 &    0.28  & 0.17  &   2 &   0.21  &  0.09   &   3\\
SMSS~J221334.13--072604.1     &    0.31  & 0.18  & 2 & 0.70 & 0.03 & 2 &    0.38 &  0.03 & 2 &    0.46  & 0.33  &   2 &   0.26  &  0.17   &   6\\
SMSS~J222349.48--114751.1     &    0.47  & 0.00  & 2 & 0.60 & 0.18 & 2 &    0.45 &  --   & 1 &    0.43  & 0.15  &   2 &   0.35$^{(1)}$ &  0.20   &   5\\
SMSS~J225336.83--270435.4     &    0.32  & 0.25  & 2 & 0.58 & 0.14 & 2 &    0.48 &  --   & 1 &    --    & --    &   - &   0.03$^{(1)}$ &  0.03   &   2\\
SMSS~J230306.22--041621.8     &    0.48  & 0.01  & 2 & 0.75 & 0.25 & 2 &    0.45 &  0.03 & 2 &    0.32  & 0.10  &   5 &   0.15  &  0.10   &   7\\
SMSS~J010839.58--285701.5     &    0.40  & 0.15  & 3 & 0.74 & 0.27 & 2 &    0.43 &  --   & 1 &    0.37  & 0.11  &   7 &   0.34  &  0.08   &   6\\
SMSS~J034249.52--284215.8     & $-$0.24  & 0.06  & 2 & 0.11 & 0.29 & 2 & $-$0.08 &  0.01 & 2 & $-$0.28  & 0.08  &   4 &$-$0.20  &  0.09   &   8\\
\hline
\end{tabular}
\end{center}
Notes:
(1) Given the lack of Fe\,{\sc II} abundances, Fe\,{\sc I} has been used for the Ti\,{\sc II} abundances relative to Fe. 
\end{table*}
%

\subsubsection{Iron-peak elements} 
A few lines were available for each of the iron-peak elements Sc, Cr,
Mn, Co, Ni and Zn (see Tab.~\ref{tab:abuFEPEAK}).   The abundances for
these elements were determined from the measured EWs except for Mn
where we synthesised the triplet at  $\approx$ 4033~\AA\ to take into
account hyperfine structure. 

\begin{sidewaystable}
\centering
\caption{Chemical abundances for Fe\,{\sc i} and Fe\,{\sc ii}, and
  Fe-peak elements Sc, Cr, Mn, Co, Ni and Zn. For each element we report
  the number (\#) of analysed spectral features and the resulting rms
  ($\sigma$). }
\label{tab:abuFEPEAK}
\scriptsize
\begin{tabular}{c ccc ccc ccc ccc cc ccc ccc ccc ccc}
\hline
ID &[Fe\,{\sc i}/H]&$\sigma$&\#& [Fe\,{\sc ii}/H]&$\sigma$&\#& [Sc/Fe]&$\sigma$&\# &[Cr\,{\sc i}/Fe]& $\sigma$&\# &[Cr\,{\sc ii}/Fe]&\#& [Mn/Fe]&$\sigma$&\#&[Co/Fe]&$\sigma$&\#& [Ni/Fe]&$\sigma$&\#& [Zn/Fe]&$\sigma$&\#\\
\hline
SMSS~J202059.17--043447.0 & $-$2.30 & 0.17 & 38 & $-$2.36 & 0.07 & 4 & \ \   0.13 & 0.35 & 4 &$-$0.19              & 0.09 & 6 &   0.13                & 1 & $-$0.25 & 0.03 & 3  &$+$0.10 & 0.16 & 6&$-$0.02                  & --   & 1 &   0.06 & 0.02 & 2 \\
SMSS~J202400.03--024445.9 & $-$1.97 & 0.19 & 36 & $-$1.81 & 0.11 & 4 & \ \   0.24 & 0.32 & 4 &$-$0.21              & 0.11 & 6 &$-$0.04   	      & 1 & $-$0.78 & 0.21 & 3  &$+$0.28 & 0.29 & 6&$-$0.22 		 & 0.01 & 2 &   0.28 & 0.11 & 2 \\
SMSS~J202601.44--033002.5 & $-$2.84 & 0.15 & 27 & $-$2.54 & 0.12 & 2 & \ \   0.00 & 0.00 & 1 & \ \            0.09 & 0.20 & 2 &     --   	      & - & $-$0.56 & 0.18 & 3  &$+$0.36 & 0.23 & 6&     --   		 & --   & - &   0.47 &   -- & 1 \\
SMSS~J202659.41--031149.2 & $-$2.94 & 0.14 & 30 & $-$2.84 & 0.17 & 2 & \ \   0.41 & 0.00 & 1 &$-$0.12   	   & 0.15 & 3 &     --   	      & - & $-$0.84 & 0.03 & 3  &$+$0.32 & 0.23 & 8&     --   		 & --   & - &     -- &   -- & - \\
SMSS~J204654.92--020409.2 & $-$1.71 & 0.14 & 43 & $-$1.61 & 0.09 & 4 & \ \   0.07 & 0.46 & 4 &$-$0.22   	   & 0.09 & 6 & \ \            0.08   & 1 & $-$0.74 & 0.08 & 3  &$+$0.20 & 0.24 & 6&$-$0.20                  & 0.10 & 8 &   0.00 & 0.01 & 2 \\
SMSS~J211657.83--012517.5 & $-$2.56 & 0.15 & 29 & $-$2.42 & 0.07 & 2 &     --     & --   & - &$-$0.04   	   & 0.16 & 4 & \ \            0.17   & 1 & $-$0.57 & 0.16 & 3  &$+$0.21 & 0.35 & 5& \ \            0.18     & --   & 1 &     -- &   -- & - \\
SMSS~J212001.71--001158.6 & $-$2.58 & 0.16 & 38 & $-$2.48 & 0.07 & 4 & \ \   0.29 & 0.23 & 2 &$-$0.23   	   & 0.11 & 5 & \ \            0.24   & 1 & $-$0.56 & 0.04 & 3  &$+$0.21 & 0.37 & 4&$-$0.17   		 & --   & 1 &   0.42 & 0.03 & 2 \\
SMSS~J212113.63--005132.2 & $-$2.75 & 0.21 & 35 & $-$2.50 & 0.13 & 2 & \ \   0.02 & 0.00 & 1 &$-$0.08   	   & 0.31 & 3 &     --   	      & - & $-$0.57 & 0.17 & 3  &$-$0.11 & 0.15 & 4&     --   	  	 & --   & - &   0.45 & 0.07 & 2 \\
SMSS~J212217.52--295552.7 & $-$2.81 & 0.14 & 10 & $-$2.62 & --   & 1 &     --     & --   & - &    --     	   & --   & - &     --   	      & - & $-$0.55 & 0.11 & 3  & --     & --   & -&     --   		 & --   & - &     -- &   -- & - \\
SMSS~J220514.10--013407.6 & $-$2.07 & 0.15 & 43 & $-$2.02 & 0.07 & 4 &   0.29     & 0.43 & 4 &$-$0.15   	   & 0.03 & 6 & \ \            0.22   & 1 & $-$0.58 & 0.21 & 3  &$+$0.45 & 0.11 & 4& \ \            0.06     & 0.14 & 6 &   0.22 & 0.07 & 2 \\
SMSS~J220535.05--004403.6 & $-$3.10 & 0.18 & 29 & $-$3.06 & --   & 1 & \ \   0.34 & 0.40 & 2 &$-$0.20   	   & 0.11 & 2 &     --   	      & - & $-$0.44 & 0.11 & 3  &$+$0.12 & 0.07 & 5&     --   		 & --   & - &     -- &   -- & - \\
SMSS~J221334.13--072604.1 & $-$2.89 & 0.15 & 24 & $-$2.80 & --   & 1 &     --     & --   & - &$-$0.10   	   & 0.13 & 2 &     --   	      & - & $-$0.69 & 0.15 & 3  &$+$0.19 & 0.29 & 7&     --   		 & --   & - &   0.59 &   -- & 1 \\
SMSS~J222349.48--114751.1 & $-$3.15 & 0.16 & 21 &     --  &   -- & - &     --     & --   & - &$-$0.28   	   & --   & 1 &     --   	      & - & $-$1.13 & 0.09 & 3  &$+$0.16 & 0.19 & 7&     --   		 & --   & - &     -- &   -- & - \\
SMSS~J225336.83--270435.4 & $-$3.26 & 0.19 &  8 &     --  &   -- & - &     --     & --   & - &    --     	   & --   & - &     --   	      & - & $-$1.05 & 0.04 & 3  &$+$0.18 & 0.36 & 4&     --   		 & --   & - &     -- &   -- & - \\
SMSS~J230306.22--041621.8 & $-$2.74 & 0.16 & 32 & $-$2.53 & --   & 1 &$-$0.14 	  & 0.00 & 1 &$-$0.17   	   & 0.03 & 2 &     --   	      & - & $-$0.54 & 0.05 & 3  &$+$0.28 & 0.12 & 7&     --   		 & --   & - &   0.38 & --   & 1 \\
SMSS~J010839.58--285701.5 & $-$2.90 & 0.13 & 33 & $-$2.91 &  --  & 1 &     --     & --   & - &$-$0.20   	   & 0.03 & 2 &     --   	      & - & $-$0.59 & 0.16 & 3  &$+$0.24 & 0.22 & 9&     --   		 & --   & - &   0.57 & 0.00 & 2 \\
SMSS~J034249.52--284215.8 & $-$2.31 & 0.13 & 42 & $-$2.21 & 0.06 & 4 &$-$0.41     & 0.00 & 1 &$-$0.31   	   & 0.11 & 6 & \ \            0.19   & 1 & $-$0.62 & 0.05 & 3  &$-$0.35 & 0.20 & 8&$-$0.17 		 & 0.10 & 3 &     -- &   -- & - \\
\hline
\end{tabular}
\end{sidewaystable}

\subsubsection{Neutron-capture elements}

\begin{figure*}
   \centering
   \includegraphics[width=0.88\textwidth,angle=0]{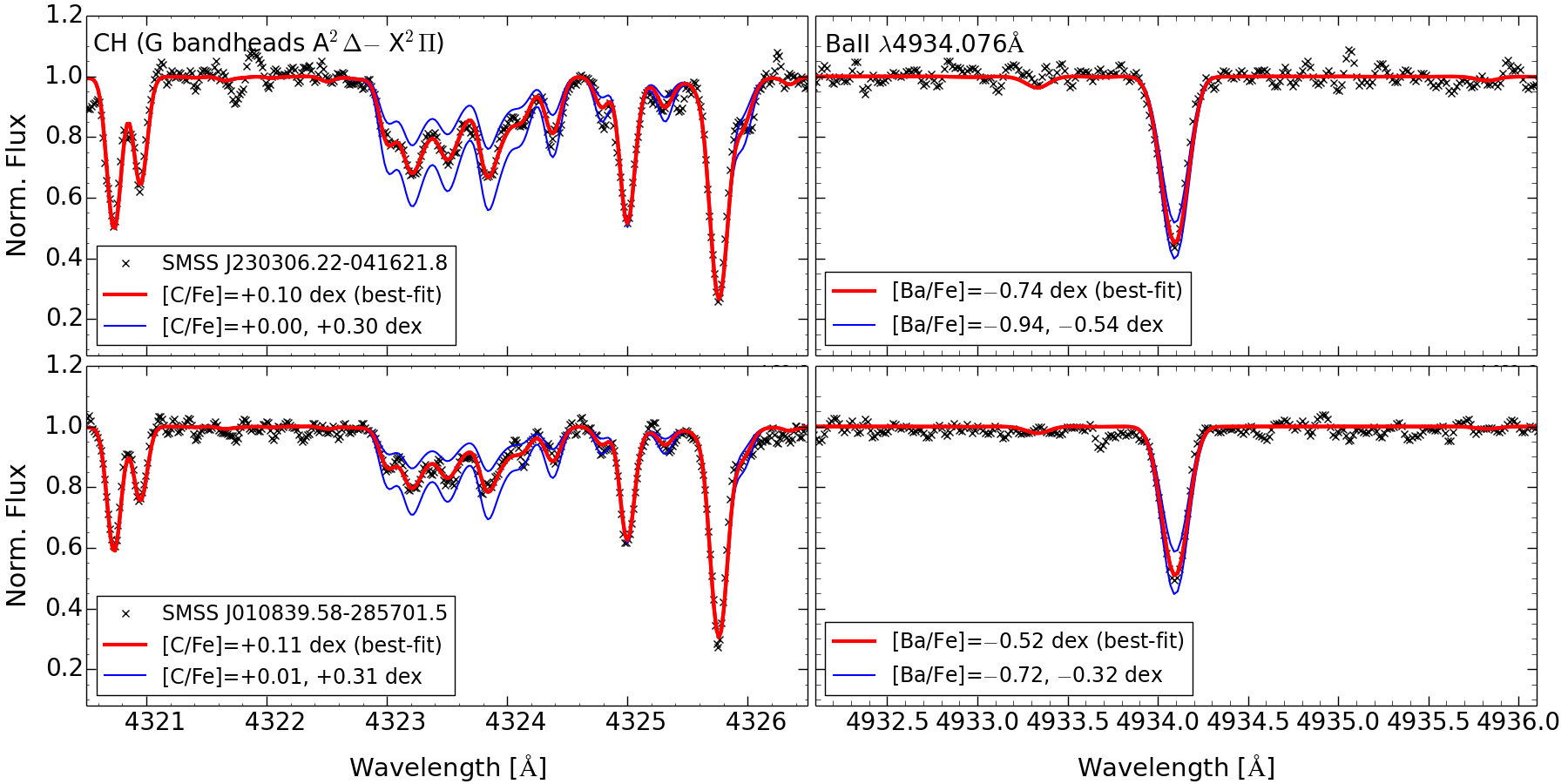}
      \caption{{\it Left panels:} synthetic spectra fits to the CH features in the vicinity of 4323~\AA, one of two spectral regions used to determine the carbon abundance.  For each star the observations are shown as black crosses, the best-fit is shown by the thick red line while the thin blue lines show [C/Fe] values 0.2 dex larger and 0.1 dex smaller than the best-fit value.  {\it Right panels:} synthetic spectra fits to the Ba\,{\sc ii} line at 4934.1~\AA, one of two (or three) lines used to determine the barium abundance.  Again the observations are shown as black crosses, the best fit is the red thick line and the thin blue lines show [Ba/Fe] values $\pm$0.2 dex about the best fit value.}
        \label{spfits}
   \end{figure*}

We derived abundances for the neutron-capture elements Sr (from the
resonance lines 4078,4215~\AA), Ba (from the resonance lines
4554,4934~\AA\ and the spectral feature 5854~\AA),
and Eu (from the resonance line 4130~\AA). 
Specifically, we employed a spectrum synthesis approach to the
analysis since hyperfine and/or isotopic splitting and/or blended
features needed to be taken into account.  For example, the spectral
features of Eu\,{\sc ii} have both significant hyperfine substructure
and isotopic splitting. For this element solar-system isotopic
fractions were assumed in the computation.   
The right panels of Fig.\ \ref{spfits} show examples of the synthetic
spectrum fits to the strong Ba\,{\sc ii} line at 4934.1~\AA. 
Our Ba abundances were computed assuming the \citet{McW98}
$r$-process isotopic composition and hyperfine splitting.     
The derived abundances are listed in Tab.~\ref{tab:abuSPROC}.

\begin{table*}
\caption{Chemical abundances for the $n$-capture elements, Sr, Ba, and
  Eu. For each element we report the number (\#) of analysed spectral
  features and the resulting rms ($\sigma$). For some stars we report
  upper limits. }
\label{tab:abuSPROC}
\begin{center}
\begin{tabular}{c ccc ccc ccc}
\hline
ID &  [Sr/Fe] & $\sigma$ &  \# & [Ba/Fe] & $\sigma$ &\# & [Eu/Fe] & $\sigma$ &\# \\
\hline
SMSS~J202059.17--043447.0 & \ \ 0.12   &  0.08      & 2    & \ \ $-$0.42   &  0.13 & 3 & $-$0.07    &  --    & 1   \\
SMSS~J202400.03--024445.9 & \ \ 0.20   &  0.29      & 2    & \ \ \ \ 0.05  &  0.04 & 3 & \ \ 0.68   &  0.03  & 2   \\
SMSS~J202601.44--033002.5 & \ \ 0.20   &  0.04      & 2    & \ \ $-$0.41   &  0.10 & 2 & $<$0.80    &  --    & -   \\
SMSS~J202659.41--031149.2 &  $-$1.94   &  0.06      & 2    & \ \ $-$1.69   &  0.00 & 1 & $<$0.10    &  --    & -   \\
SMSS~J204654.92--020409.2 & \ \ 0.03   &  0.00      & 2    & \ \ $-$0.13   &  0.16 & 3 & \ \ 0.35   &  0.19  &  2  \\
SMSS~J211657.83--012517.5 & $-$0.18   &  0.21      & 2    & \ \ $-$0.41   &  0.09 & 3 & \ \ 0.28   &  --    &  1  \\
SMSS~J212001.71--001158.6 & $-$0.41   &  0.07      & 2    & \ \ $-$0.82   &  0.15 & 3 & $-$0.28    &  --    & 1   \\
SMSS~J212113.63--005132.2 & $-$0.51   &  0.01      & 2    & \ \ $-$1.00   &  0.02 & 2 & $<$0.20    &  --    &  -  \\
SMSS~J212217.52--295552.7 & $-$1.15   &  0.05      & 2    &    $<-$0.90   &  0.10 & 2 & $<$0.80    &  --    &  -  \\
SMSS~J220514.10--013407.6 & \ \ 0.03   &  0.07      & 2    & \ \ $-$0.38   &  0.10 & 3 & \ \ 0.05   &  --    &  1  \\
SMSS~J220535.05--004403.6 & $-$0.82   &  0.03      & 2    & \ \ $-$1.45   &  0.09 & 2 & $<$0.30    &  --    &  -  \\
SMSS~J221334.13--072604.1 & $-$0.36   &  0.04      & 2    & \ \ $-$1.00   &  0.09 & 2 & $<$0.30    &  --    &  -  \\
SMSS~J222349.48--114751.1 &  $-$2.00   &  0.10      & 2    & \ \ $-$1.63   &  0.02 & 2 & $<$0.20    &  --    &  -  \\
SMSS~J225336.83--270435.4 & $-$0.61   &  0.10      & 2    & \ \ $-$1.12   &  0.15 & 2 & $<$0.90    &  --    &  -  \\
SMSS~J230306.22--041621.8 &  $-$0.01   &  0.01      & 2    & \ \ $-$0.58   &  0.10 & 3 & $<$0.30    &  --    &  -  \\
SMSS~J010839.58--285701.5 & \ \ 0.07   &  0.01      & 2    & \ \ $-$0.41   &  0.11 & 2 & $<$0.30    &  --    &  -  \\
SMSS~J034249.52--284215.8 &  $-$0.10   &  0.01      & 2    & \ \ $-$0.14   &  0.10 & 3 & \ \ 0.55   &  0.01  &  2  \\
\hline
\end{tabular}
\end{center}
\end{table*}

%

\subsubsection{Abundance errors}

Estimates of the uncertainties in the chemical abundances due to
errors in the atmospheric parameteres have been
obtained by rerunning the abundances, one at a time, varying
\teff/\logg/[m/H]/\vmicro\  by
$\pm$100~K/$\pm$0.40/$\pm$0.30/$\pm$0.40~\kmsec, assuming that the
errors are symmetric for positive and negative changes. 
The uncertainties used in \teff, \logg\ and [m/H] are reasonable, as
suggested by the comparison with the spectrophotometric fits to the
2.3m low-resolution spectra and stars in common with \citet{HJ15} (see
Sect.\ \ref{sec:atm}).
As internal errors in \vmicro, we conservatively adopt $\pm$0.40~\kmsec.
The variations in chemical abundances for each element are listed in
Table\ \ref{tab:errors}. 

To obtain the total error estimates we follow the approach by
\citet{JEN10} and \citet{DY13}. For each element, we replace the
r.m.s ($\sigma$) in Tables\
\ref{tab:abuCNNa}, \ref{tab:abuALPHA}, \ref{tab:abuFEPEAK} and \ref{tab:abuSPROC} by the
maximum($\sigma$, 0.20), where the second term is what would be
expected for a set of $N$ lines ($N_{\rm lines}$) with a dispersion of 0.20~dex (a
conservative value for the abundance dispersion of Fe~{\sc i} lines as
listed in Table\ \ref{tab:abuFEPEAK}). Then, we derive max($\sigma$,
0.20)/$\sqrt{N_{\rm lines}}$. Typical values obtained for each
element are listed in column (6) of Table\ \ref{tab:errors}.
The total error is obtained by quadratically adding this random error
with the uncertainties introduced by atmospheric parameters. 
For Sr and Ba we conservatively adopt an uncertainty of 0.30~dex,
considering that the abundances for these elements mostly come from
strong resonance lines.
Finally, we note that this 1D LTE analysis is subject to abundance
uncertainties from three-dimensional (3D) and non-LTE effects \citep{MA05}. 

\begin{table*}
\caption{Sensitivity of derived abundances to the uncertainties in
  atmospheric parameters, the limilted S/N ($\sigma_{\rm S/N}$) and
  the total error due to these contributions ($\sigma_{\rm
    tot}$).}\label{tab:errors} 
\begin{center}
\begin{tabular}{lcccccc}
\hline
  & $\Delta$\teff & $\Delta$\logg & $\Delta$\vmicro  & $\Delta$[A/H] & $\sigma_{\rm S/N}$ & $\sigma_{\rm total}$\\
  & $\pm$100~K    & $\pm$0.40     & $\pm$0.40~\kmsec & 0.30~dex      &                   &                   \\
\hline
$\rm {[C/Fe]}$            &$\pm$0.09  &$\mp$0.10  & $\pm$0.05  & $\pm$0.11  & 0.14  & 0.23  \\
$\rm {[N/Fe]}$            &$\pm$0.16  &$\mp$0.08  & $\pm$0.06  & $\mp$0.20  & 0.14  & 0.31  \\
$\rm {[Na/Fe]}$           &$\pm$0.01  &$\mp$0.05  & $\mp$0.10  & $\mp$0.01  & 0.14  & 0.18  \\
$\rm {[Mg/Fe]}$           &$\mp$0.06  &$\mp$0.01  & $\pm$0.02  & $\pm$0.01  & 0.14  & 0.15  \\
$\rm {[Al/Fe]}$           &$\pm$0.01  &$\mp$0.12  & $\mp$0.10  & $\pm$0.03  & 0.35  & 0.38  \\
$\rm {[Si/Fe]}$           &$\mp$0.01  &$\mp$0.05  & $\mp$0.03  & $\pm$0.01  & 0.33  & 0.34  \\
$\rm {[Ca/Fe]}$           &$\mp$0.05  &$\pm$0.01  & $\pm$0.04  & $\pm$0.01  & 0.14  & 0.15  \\
$[$Ti\,{\sc i}/Fe$]$      &$\pm$0.02  &$\pm$0.00  & $\pm$0.05  & $\pm$0.00  & 0.10  & 0.11  \\
$[$Ti\,{\sc ii}/Fe$]$     &$\pm$0.04  &$\mp$0.01  & $\mp$0.08  & $\pm$0.00  & 0.08  & 0.12  \\
$[$Fe\,{\sc i}/H$]$       &$\pm$0.12  &$\mp$0.03  & $\mp$0.08  & $\mp$0.02  & 0.04  & 0.15  \\
$[$Fe\,{\sc ii}/H$]$      &$\mp$0.00  &$\pm$0.12  & $\mp$0.03  & $\pm$0.02  & 0.14  & 0.19  \\
$\rm {[Sc/Fe]}$           &$\pm$0.02  &$\mp$0.04  & $\pm$0.01  & $\pm$0.00  & 0.33  & 0.33  \\
$[$Cr\,{\sc i}/Fe$]$      &$\mp$0.01  &$\pm$0.00  & $\pm$0.05  & $\pm$0.01  & 0.16  & 0.17  \\
$[$Cr\,{\sc ii}/Fe$]$     &$\mp$0.01  &$\mp$0.07  & $\pm$0.02  & $\mp$0.01  & 0.20  & 0.21  \\
$\rm {[Mn/Fe]}$           &$\pm$0.06  &$\mp$0.07  & $\mp$0.08  & $\mp$0.13  & 0.12  & 0.21  \\
$\rm {[Co/Fe]}$           &$\pm$0.02  &$\mp$0.03  & $\mp$0.07  & $\mp$0.01  & 0.17  & 0.19  \\
$\rm {[Ni/Fe]}$           &$\mp$0.06  &$\pm$0.03  & $\pm$0.08  & $\pm$0.01  & 0.12  & 0.16  \\
$\rm {[Zn/Fe]}$           &$\mp$0.09  &$\pm$0.07  & $\pm$0.07  & $\pm$0.02  & 0.14  & 0.19  \\
$\rm {[Sr/Fe]}$           &$\pm$0.07  &$\mp$0.06  & $\mp$0.22  & $\mp$0.04  & 0.14  & 0.30  \\
$\rm {[Ba/Fe]}$           &$\pm$0.07  &$\mp$0.02  & $\mp$0.05  & $\mp$0.02  & 0.14  & 0.30  \\
$\rm {[Eu/Fe]}$           &$\pm$0.05  &$\pm$0.01  & $\pm$0.03  & $\pm$0.06  & 0.20  & 0.22  \\
\hline
\end{tabular}
\end{center}
\end{table*}

\section{Results} \label{Sect4}

\begin{figure}
   \centering
   \includegraphics[width=0.48\textwidth,angle=0]{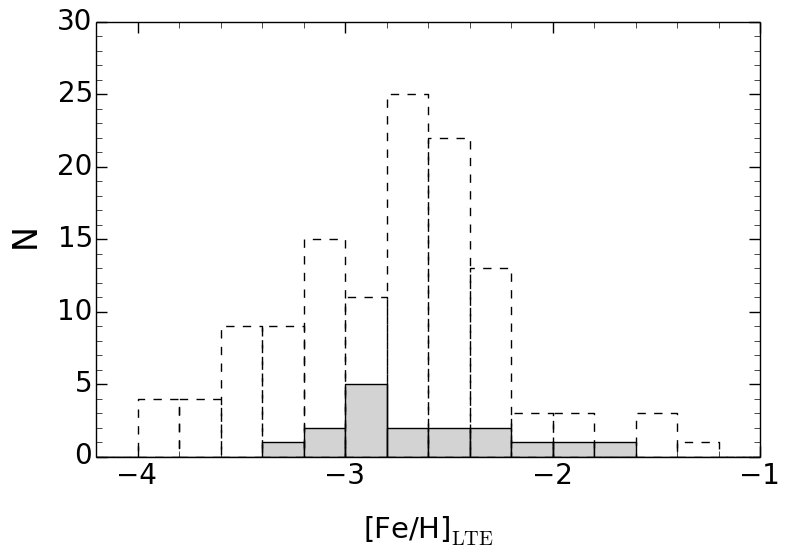}
      \caption{Distribution of the [Fe/H]$_{\mathrm {LTE}}$ abundances of our
        analysed stars (filled-grey histogram) and of the larger sample
        analysed in \citet{HJ15} (dashed-empty histogram).  
      }
        \label{fig:fe}
   \end{figure}

The distribution of [Fe/H]$_{\mathrm {LTE}}$ for the sample of 17 commissioning-era
SkyMapper EMP candidates observed at Keck is shown in
Fig.~\ref{fig:fe}, where it is compared with the [Fe/H]$_{\mathrm {LTE}}$
distribution for the larger sample of 122 commissioning-era SkyMapper
EMP candidates observed at Magellan and analysed in \citet{HJ15}. In
the \citet{HJ15} sample one-third of the stars have [Fe/H]$_{\mathrm
  {LTE}} < -$3.0, while 43\% have [Fe/H]$_{\mathrm {LTE}}$ $<
-$2.8~dex. Given the smaller size, the current sample is fully consistent with these fractions as
$\sim$20\% (3/17) of the stars are found here to have [Fe/H]$_{\mathrm {LTE}}$
$< -$3.0~dex and 47\% (8/17) have [Fe/H]$_{\mathrm {LTE}}$ $< -$2.8~dex.  
The SkyMapper photometric selection technique is therefore clearly
quite efficient in selecting metal-poor stars.  
Indeed, as noted above, using the SkyMapper DR1.1 photometry,
$\sim$40\% of the candidates that fall within the selection window
shown in Fig.\ \ref{fig:mi_vs_gi}, have [Fe/H]$_{2.3m}$$\leq -$2.75~dex. 
As a comparison, the similar northern hemisphere photometric survey
for EMP stars, {\it Pristine}, finds that $\sim$24\% of candidates
photometrically selected to have [Fe/H]$< -$3.0 have spectroscopically determined
abundances below [Fe/H]=$-$3.0~dex \citep{ES17}.  As in \citet{HJ15},
we caution against using the commissioning-era results to constrain
the metallicity distribution function at low abundances, as the
selection biases cannot be reliably established.  Future papers based
on a much larger sample of stars selected from SkyMapper DR1.1
photometry and observed at low resolution, coupled with an extensive
follow-up investigation with Magellan, will, however, address this
issue.

In the following subsections we consider the abundance trends among
and between elements of different nucleosynthetic groups.  We use as
our comparison samples those of \citet{HJ15} and the giant stars in
the compilation of \citet{DY13}, noting that the parameter
determination approaches and the line-lists in those works are not
identical to those used here so that the possibility of systematic
differences cannot be ruled out. Unless otherwise noted all abundances
and abundance ratios are 1D LTE values.

%
%

\subsection{Light elements}\label{sec:light}

\subsubsection{Carbon}

As a star ascends the red giant branch, 
the envelope expands inwards, reaching layers affected by CN-cycling,
a consequence of which is a reduction of the carbon abundance in the
surface layers (and an increase in the surface abundance of N). 
Since we are interested in the carbon abundance
at the star's birth, the so-called `natal' abundance, our measured
carbon abundances need to be corrected for the effects of this
evolutionary mixing. The evolutionary mixing corrections depend on
\teff, \logg\ and [Fe/H] and have been discussed in detail in
\citet{VP14}. 
Dr.\ V.\ Placco (Placco, 2018, {\it priv.\ comm.}) kindly generated
the appropriate corrections to our observed carbon abundances by
assuming a natal [N/Fe] = +0.0 and applying the \citet{VP14}
procedure.  Table \ref{tab:abuCNNa} lists the observed [C/Fe] values
and the correction for evolutionary mixing: the estimated `natal'
[C/Fe] is formed by adding the correction to the observed value. 

\begin{figure*}
\centering
\includegraphics[width=1.0\textwidth,angle=0.]{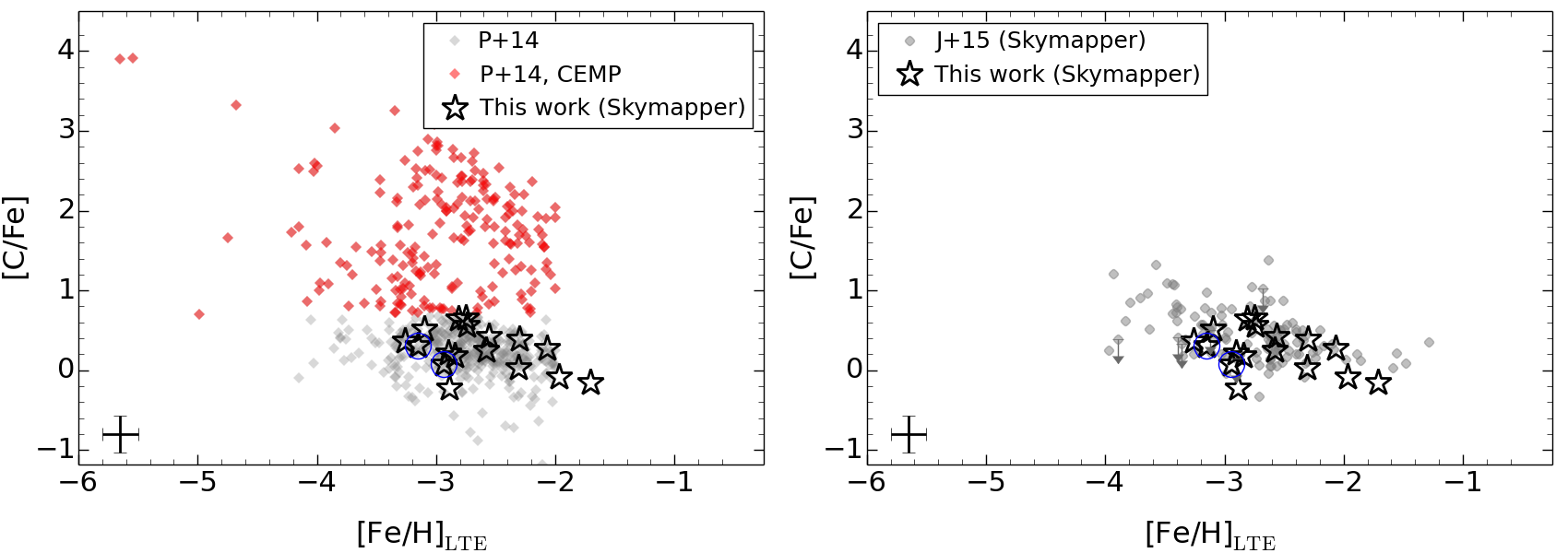}
\caption{[C/Fe], corrected for evolutionary mixing effects, as a
  function of [Fe/H]$_{\mathrm {LTE}}$ for our sample of stars,
  which are shown as black open 5-pointed star symbols. The left panel
  shows the comparison of our observed values with the compilation of
  \citet{VP14}, shown as grey and red filled diamonds with the latter
  marking CEMP stars. In the right panel we compare our
  evolutionary-mixing corrected values with those of \citet{HJ15},
  plotted as grey filled circles, that are also corrected for
  evolutionary-mixing effects. The two stars indicated with blue open
  circles are the stars with low neutron-capture elements, as will be
  discussed in Section~\ref{sec:srba}.}  
\label{fig:carbon}
\end{figure*}

In Figure~\ref{fig:carbon} we show in the left panel a comparison of
our observed [C/Fe] values with those listed in \citet{VP14}, which
are corrected for evolutionary-mixing effects.  The right panel shows
the comparison of our [C/Fe] values, after applying the evolutionary
mixing corrections, with the evolutionary mixing corrected values of
\citet{HJ15}. 
\citet{VP14} have demonstrated that the fraction of carbon-enhanced
metal-poor (CEMP) stars, defined as stars possessing [C/Fe] $\geq
+$0.7~dex, increases with decreasing metallicity with CEMP stars
dominant below [Fe/H] $\leq$$-$4.0 \citep[see also][]{Yoon18}.  The
\citet{VP14} CEMP frequencies (e.g., $\sim$40\% for
[Fe/H]$\leq$$-$3.0) would suggest that our sample of three stars with
[Fe/H]$_{\mathrm {LTE}} \leq -$3.0 should contain one CEMP-star, whereas there are
none.  While the statistical weight of the lack of CEMP-stars compared
to the number expected is not high, inspection of Fig.\
\ref{fig:carbon} reveals that none of our sample of 17 stars has a
[C/Fe] value that would cause it to be classified as CEMP-star: the
highest evolutionary corrected [C/Fe] values are 0.66 dex for  
SMSS~J212113.63--005132.2 ([Fe/H]$_{\mathrm {LTE}}$=$-$2.74) and 0.64 dex
for SMSS~J212217.52--295552.7 ([Fe/H]$_{\mathrm {LTE}}$=$-$2.81).   
In general our evolutionary mixing corrected [C/Fe] values are
completely consistent with those from the larger sample of
\citet{HJ15}.  \citet{HJ15} discussed the frequency of CEMP-stars in
their sample and concluded that it is comparable with that of
\citet{VP14} although, as is evident in Fig.\ \ref{fig:carbon}, the
\citet{HJ15} sample lacks CEMP-stars with [C/Fe] significantly above
1.0.

The most likely explanation lies in the selection of EMP candidates
from the SkyMapper photometry.  As discussed in  
\citet{GDaC19}, the strong CH-bands in the spectrum of a CEMP-star can
depress the flux in the SkyMapper {\it v}-filter sufficiently that the
inferred metallicity index mimics a more metal-rich star, and thus
decreases the probability it will be selected for low resolution
spectroscopic follow-up.
Nevertheless the commissioning survey did result in the discovery of
the most iron-poor star currently known, a star that is extremely
C-rich \citep{SK14, MSB15, TN17}. 
Evidently at sufficiently low overall abundance the contaminating
carbon features in the $v$ band weaken enough that selection as a
photometric candidate again becomes possible.   

\subsubsection{Nitrogen, sodium and aluminum}

\begin{figure*}
\centering
\includegraphics[width=0.9\textwidth,angle=0.]{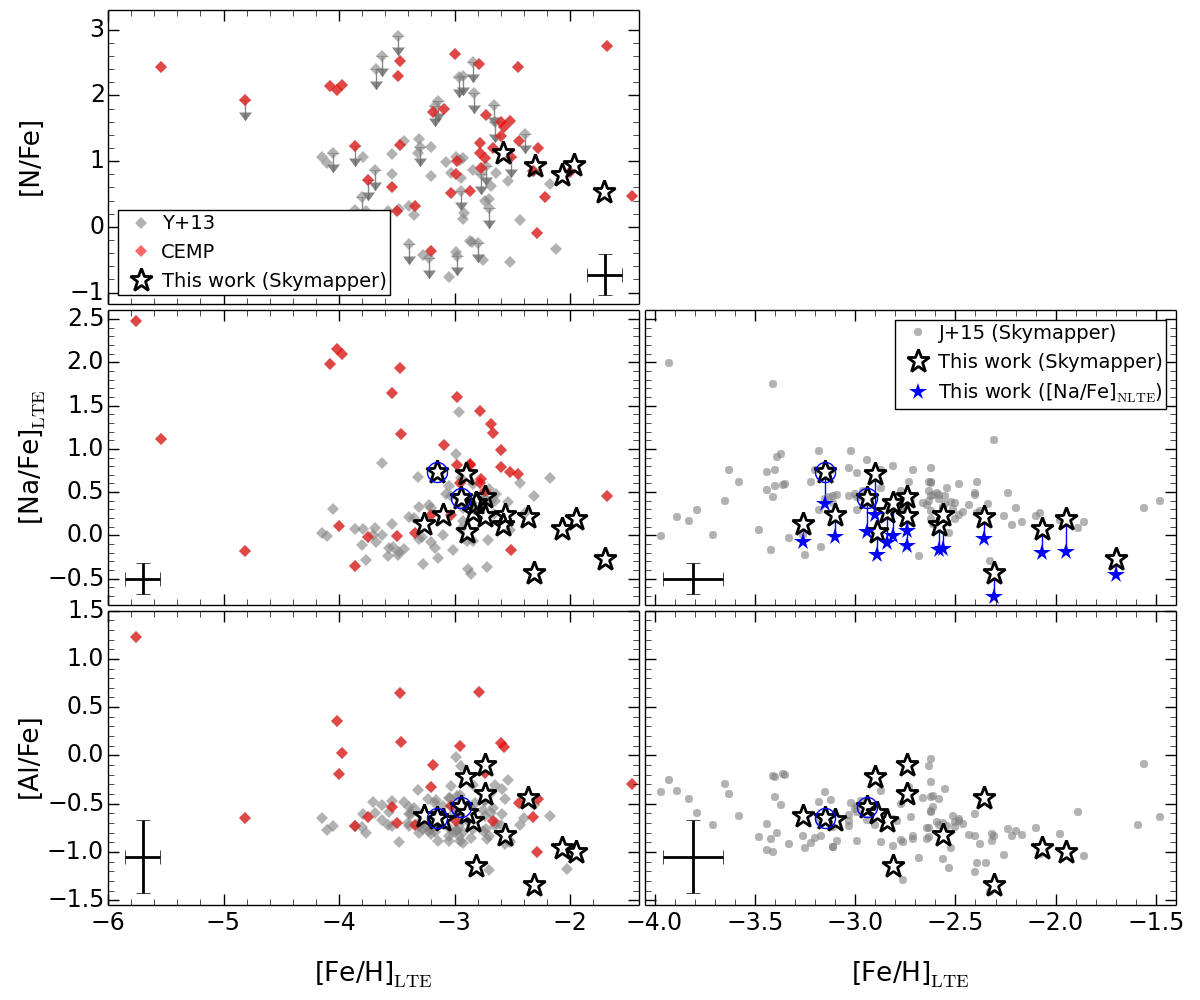}
\caption{
Nitrogen, sodium and aluminum abundance ratios, relative to iron, as a function
of [Fe/H]$_{\mathrm {LTE}}$. The [N/Fe] abundances are compared
with the sample of Yong et al.\,(2013) in the upper-left panel where
upper limits on the [N/Fe] values are shown as downward-pointing
arrows. The LTE [Na/Fe] abundance ratios derived here are compared
with the Yong et al.\,(2013) values in the middle-left panel (symbols
as for the upper-left panel), and with the Jacobson et al.\,(2015)
sample in the middle-right panel. In the middle-right panel we also show, as
blue-filled stars, our [Na/Fe] values corrected for NLTE effects.
These lie at lower values and are connected to the corresponding LTE
points by thin blue lines. Aluminum abundances are compared with
  the Yong et al.\,(2013) and the Jacobson et al.\,(2015) in the
  lower-left and lower-right panels, respectively. The two stars indicated with blue open
  circles are the stars with low neutron-capture elements, as will be
  discussed in Section~\ref{sec:srba}.}
\label{fig:light}
\end{figure*}

The nitrogen, sodium and aluminum abundance ratios with respect to iron for our
sample are shown in Fig.~\ref{fig:light} as a function of [Fe/H]$_{\mathrm {LTE}}$.
Because of low S/N at the wavelength of the 
CN-bands in many of the spectra, [N/Fe] values could be determined
only for five stars in our sample.  The values, which lie between 0.5
and $\sim$1.0, and which are listed in Table~\ref{tab:abuCNNa}, are
nevertheless consistent with the midpoint of the substantial range of
[N/Fe] values found in the sample of \citet{DY13}.   

For sodium, while noting that the NLTE corrections would result in
lower abundance ratios, we have plotted the LTE abundance ratios to
facilitate comparison with the \citet{DY13} and \citet{HJ15}
samples. In the comparison plot with \citet{HJ15} we have also plotted
our NLTE-corrected values of Na \citep{KL11}, to highlight the general lower
abundances for this element that would be obtained with a proper NLTE
analysis. It is clear from the panels of Fig.~\ref{fig:light}, that
our results for [Na/Fe] are generally consistent with those of the
earlier studies. The one possible exception is the star
SMSS~J034249.52--284215.8 which has [Na/Fe]$_{\mathrm {LTE}}$ of
$-$0.44 and [Fe/H]$_{\mathrm {LTE}}$=$-$2.31~dex.  Such a low [Na/Fe] is reminiscent of the low
[Na/Fe] values seen in red giant members of dwarf Spheroidal galaxies
\citep[e.g.][and references therein]{DG05,JEN17,H17}.  The low value
for [Na/Fe] found here is consistent with that listed by \citet{HJ15}:
[Na/Fe]$_{\mathrm {LTE}}$=$-$0.29, the lowest [Na/Fe]$_{\mathrm {LTE}}$ in their entire
sample. We give both the LTE and NLTE [Na/Fe] values for our stars in
Table~\ref{tab:abuCNNa}.  

Aluminum abundance ratios of our sample are comparable to both
those of \citet{DY13} and \citet{HJ15} (lower panels of Fig.~\ref{fig:light}).
We note that the uncertainties associated with our [Al/Fe] values are
large due to the relatively low S/N of our spectra, especially below
4000\AA. As for [Na/Fe] we expect the application of NLTE corrections to
generate systematic offsets in the [Al/Fe]$_{\mathrm {LTE}}$ values;
such corrections can be as large as $+$0.65 dex \citep{BG97}. As
discussed in previous work, such higher NLTE [Al/Fe] abundances would
be more consistent with predictions of chemical evolution models
\citep[e.g.][]{CK06}.

%
%

%
%

\subsection{$\alpha$-elements}

\begin{figure*}
\centering
\includegraphics[width=1\textwidth]{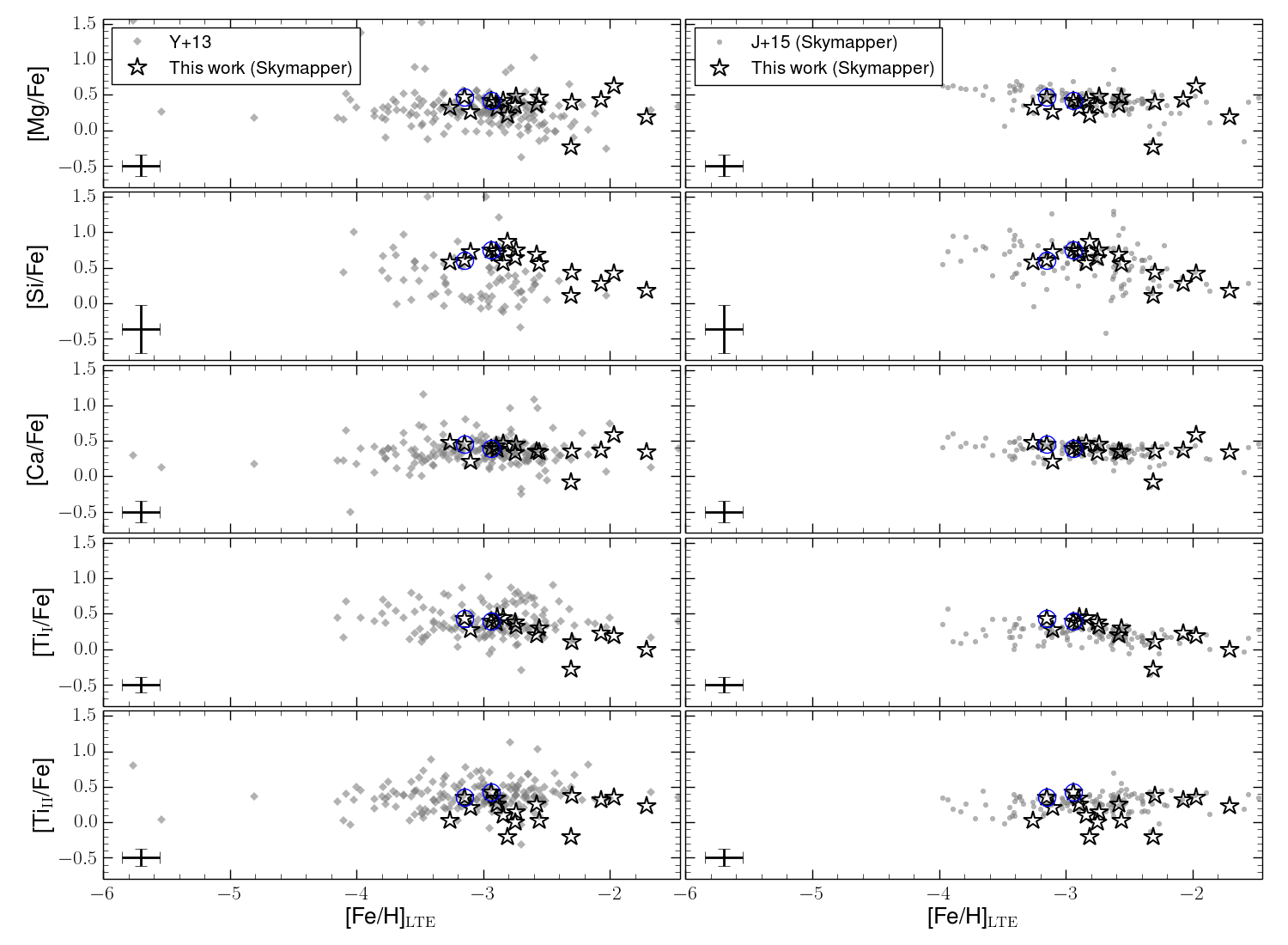}
\caption{Chemical abundance ratios with respect to Fe for the
  $\alpha$-elements measured in this study as a function of
  [Fe/H]$_{\mathrm {LTE}}$. The 5-point star symbols are the stars 
  in the current sample while the grey solid circles are stars from
  \citet[][left panels]{DY13} and \citet[][right panels]{HJ15}. 
  The two stars indicated with blue open circles are the stars with
  low neutron-capture elements, as will be discussed in
  Section~\ref{sec:srba}.}     
\label{fig:alpha}
\end{figure*}

The individual $\alpha$-element (Mg, Si, Ca, Ti\,{\sc i}, Ti\,{\sc ii})
abundances for our sample are displayed as a function  
of [Fe/H]$_{\mathrm {LTE}}$ in Fig.~\ref{fig:alpha} and listed in
Table~\ref{tab:abuALPHA}. With the exception of one star, all our
stars are $\alpha$-enhanced and their location in the [element/Fe] panels is
fully consistent with the larger comparison samples of \citet{HJ15}
and \citet{DY13}. 

The one star that does not show any $\alpha$-enhancement is the star
SMSS~J034249.52--284215.8 which was identified as a ``Fe-enhanced''
star in \citet[][specifically \S5.1]{HJ15}.  For this star we find
([Mg/Fe], [Si/Fe], [Ca/Fe], [Ti/Fe]\,{\sc i} and [Ti/Fe]\,{\sc ii}) values of
($-$0.24, $+$0.11, $-$0.08, $-$0.28, $-$0.20), values that are fully consistent
with those of \citet{HJ15}, which are ($-$0.17, $+$0.14, $-$0.16, $-$0.37,
$-$0.13).  We find also that the other elements analysed in this star
generally have sub-solar ratios, again consistent with \citet{HJ15}.
We note that in Section~\ref{sec:atm} we have used
$\alpha$-enhanced isochrones for all the stars. A solar-scaled
[$\alpha$/Fe] isochrone, more appropriate for this star, results in
a lower \logg\ by $\sim$0.10~dex, which does not significantly affect the
derived abundances relative to Fe (see Table\ \ref{tab:errors}). 

Discussion of the possible origin(s) of this star is given in
\citet{HJ15}.  We only note that as mentioned above, the low [Na/Fe]
for this star, plus its ``alpha-poor'' nature, is reminiscent of
abundance ratios seen in dSph stars.   
The kinematics of the star are not unusual in comparison with
those for the rest of the sample. This star also has the lowest
[Al/Fe] in both our sample and that of \citet{HJ15}.

%
%

\subsection{Fe-peak elements}

\begin{figure*}
\centering
\includegraphics[width=1.0\textwidth,angle=0.]{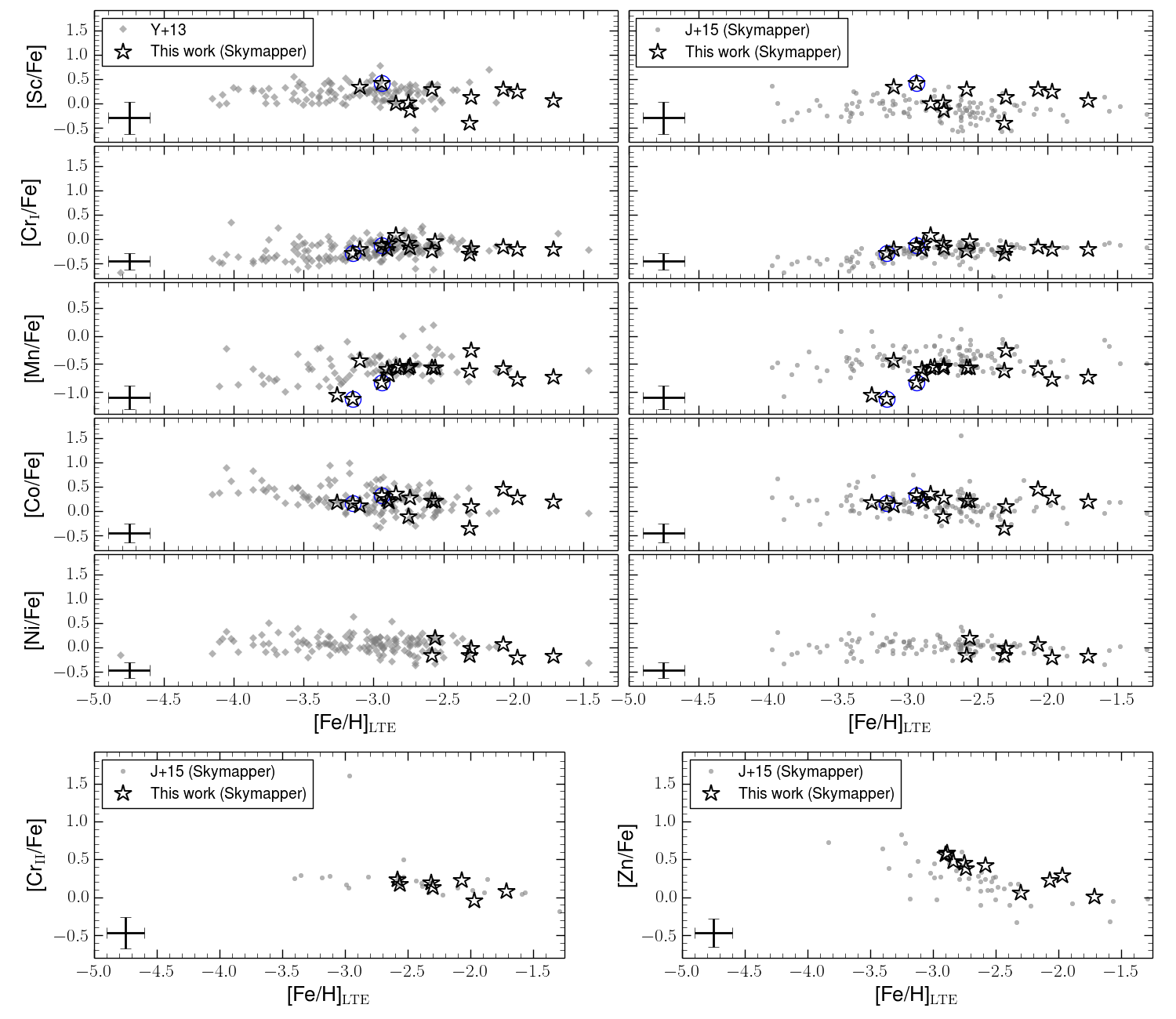}
\caption{Chemical abundance ratios with respect to Fe for the
  iron-peak elements as a function of [Fe/H]$_{\mathrm {LTE}}$.  
Symbols are as in Fig.~\ref{fig:alpha}.  The left panels show the
comparison with the results of \citet{DY13} while the right panels
show the comparison with \citet{HJ15}.      
The two stars indicated with blue open
circles are the stars with low neutron-capture elements, as will be
discussed in Section~\ref{sec:srba}.} 
\label{fig:ironp}
\end{figure*}

In Figure~\ref{fig:ironp} we show our results for the abundance ratios
with respect to iron for the iron-peak elements Sc, Cr\,{\sc i},
Cr\,{\sc ii}, Mn, Co, Ni and Zn as a function of [Fe/H]$_{\mathrm {LTE}}$.   
The values are listed in Table \ref{tab:abuFEPEAK} along with both the
number of spectral features analysed and the standard deviations
($\sigma$). Also shown in the panels are the equivalent data, where
available, for the stars in the comparison samples of \citet{DY13} and
\citet{HJ15}.  Although our sample is not large compared to the
others, it is evident from the figure that our results are consistent
with the abundance ratio trends seen in the comparison samples.  There
is, however, a suggestion that the Keck data presented here have some
systematic differences relative to the comparison samples.  
For example, although the Keck stars show the same rate of increase in
[Zn/Fe] with decreasing  
[Fe/H]$_{\mathrm {LTE}}$ as the stars in the \citet{HJ15} sample, there might be
an offset in that the current sample have [Zn/Fe] abundance ratios
$\sim$0.2~dex higher than the \cite{HJ15} values at similar [Fe/H].
The star observed here that is in common with \citet{HJ15}, is
consistent with this offset.  

%
%

\subsection{$n$-capture elements}

\subsubsection{Strontium and Barium}\label{sec:srba}

Among the $n$-capture elements, those which could be analysed in the
spectra of the majority of the stars observed here are Sr and Ba.  
As regards the $s$-process, Sr is a first $s$-process peak element
while Ba occurs in the second $s$-process peak. Both can be generated
by the $^{22}$Ne or the $^{13}$C neutron source depending on the neutron
exposure. These elements can also have $r$-process contributions and
thus the relative abundances of these elements in metal-poor stars can
provide information on nucleosynthetic processes at early times.

The abundance ratios [Sr/Fe] and [Ba/Fe] for the stars in our sample
are shown as a function of [Fe/H]$_{\mathrm {LTE}}$ in the upper and middle panels of
Fig.\ {\ref{fig:sproc} and are listed in Table \ref{tab:abuSPROC}.  We
  note first that none of our stars show high ($>$1 dex) abundance
  ratios for these elements, i.e., none can be classified as
  $s$-enhanced stars.  This is consistent with the lack of CEMP-stars
  in our sample, as discussed in \S \ref{sec:light}.  As is also
  apparent in the panels of Fig.\ \ref{fig:sproc}, our results are
  generally consistent with those of \citet{DY13}, which include
  CEMP-$s$ stars, and \citet{HJ15}.  There is some indication that
  perhaps our [Sr/Fe] values are systemically lower than those of
  \citet{HJ15}, by approximately 0.3~dex, which is however within our
  observational uncertainties (see Tab.~\ref{tab:errors}). 

It is well-known that as overall abundance decreases, the dispersion
in the abundance ratios for the $n$-capture elements relative to iron
increases markedly \citep[e.g.][and references therein]{McW95,FN15},
undoubtedly reflecting variations in the relative contributions of the
numerous nucleosynthetic origins for these elements.  This is
illustrated in the lower panels of Fig.\ \ref{fig:sproc} where we show
the [Sr/Ba] ratio as a function of [Ba/Fe], including CEMP-$s$ stars.
Concentrating on the stars without $s$-enhancements, i.e., those with
[Ba/Fe] $\leq$ 0.0 approximately, we see that the range in [Sr/Ba]
increases substantially as [Ba/Fe] decreases reaching almost two
orders of magnitude at the lowest [Ba/Fe] values.  The data suggest
that the upper limit on [Sr/Ba] increases as [Ba/Fe] decreases,
whereas the lower limit appears approximately constant with decreasing
[Ba/Fe].  The solar system $r$-process pattern has [Sr/Ba]=$-$0.5~dex,
though lower values are seen in some ultra-faint dwarf galaxy member
stars \citep[see][and references therein]{FN15} and do occur in both
the \citet{DY13} and \citet{HJ15} datasets.   

\citet{CS17} found a star with [Ba/Fe]$\sim-$3, with typical main
$r$-process abundance patterns, a signature of the universality of the
$r$-process. 
Six stars with very low [Ba/Fe] ([Ba/Fe]$<-$1.5) but which show a
range of [Sr/Ba] of $\sim$2~dex, were identified by \citet{HJ15}.
One of these stars has [Sr/Ba] $\approx -$0.5, i.e., the solar system
$r$-process value, with an upper limit on [Eu/Fe] of $\sim$0.4~dex.
It is therefore strongly depleted in $n$-capture elements.  We have
identified two similar stars in our sample: SMSS~J202659.41--031149.2
for which [Fe/H]$_{\mathrm {LTE}}$=$-$2.94, \mbox{[Ba/Fe]=$-$1.69} and
[Sr/Ba]=$-$0.25, and SMSS~J222349.48--114751.1 for which the
corresponding values are $-$3.15, $-$1.63 and $-$0.37 dex.  Neither star has a
detectable Eu\,{\sc ii} line at 4129~\AA\/ yielding an
approximate upper limit of [Eu/Fe] $\approx$ 0.10-0.20~dex.  
Detailed abundances for other $n$-capture elements for these stars
would provide important information on $n$-capture nucleosynthesis
processes at early times, e.g., the weak $r$-process versus the main
$r$-process \citep[e.g.][]{IR13, Li15}. 

%
\begin{figure*}
\centering
\includegraphics[width=1\textwidth,angle=0.]{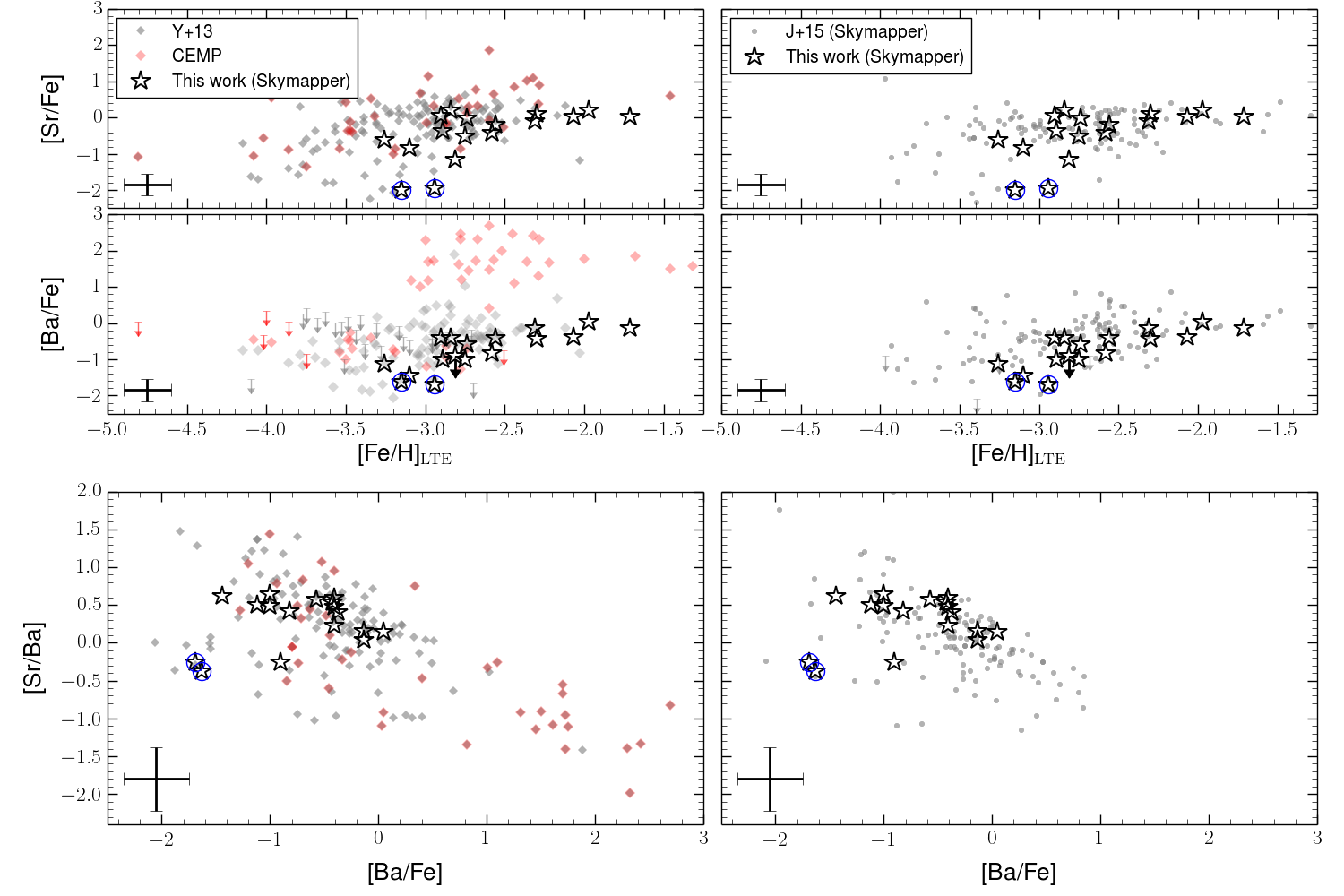}
\caption{{\it Upper panels:} Chemical abundance ratios with respect to
  Fe for the neutron-capture elements Sr and Ba as a function of
  [Fe/H]$_{\mathrm {LTE}}$. {\it Lower panels:} [Sr/Ba] as function of
  [Ba/Fe]. We compare our results with data from \citet{DY13} in the
  left panels and with \citet{HJ15} in the right panels.  Symbols are
  as in Fig.~\ref{fig:carbon}.  
The two stars indicated with blue open circles are the stars with low
neutron-capture elements.} 
\label{fig:sproc}
\end{figure*}
%

\subsubsection{Europium}

Europium is predominantly synthesized by the $r$-process
\citep[e.g.][]{CS08} and as such, the [Eu/Fe] abundance ratio is used
to identify $r$-processed enhanced stars: $r$-II stars have [Eu/Fe]
$\geq$ +1.0 while the more moderately enhanced $r$-I stars have
0.3$\leq$[Eu/Fe]$\leq$1.0~dex.  Both types have [Ba/Eu] $<$ 0
\citep{PB05}.  We have measured Eu abundances for as many of the stars
in our sample as possible, and derived upper limits for the others.
The results are shown in the upper panel of Fig.\ \ref{fig:eu} where we compare our
results with those of \citet{HJ15} (we note that \citet{DY13} did not
determine Eu abundances).  The agreement is reasonable.  The [Eu/Fe]
determinations are also given in Table \ref{tab:abuSPROC}. Overall,
the scatter in the [Eu/Fe] values is comparable to that seen in
\citet{HJ15} and to that in the literature compilation of
\citet{AF10}.  One star in our sample, SMSS~J202400.03--024445.9 is a
probable $r$-I star -- for this star, which has [Fe/H]$_{\mathrm {LTE}}$=$-$1.95, we
find [Eu/Fe]=$+$0.68 and [Ba/Eu]=$-$0.63~dex. Star
SMSS~J034249.52--284215.8, which we have already drawn attention to
because of its low values of [$\alpha$/Fe] and [Na/Fe], and which is the
``Fe-enhanced'' star discussed in \citet{HJ15}, is also a candidate
$r$-I star. It has [Fe/H]$_{\mathrm {LTE}}$=$-$2.31, [Eu/Fe]=$+$0.55 and
[Ba/Eu]=$-$0.67 dex. Star SMSS~J204654.92--020409.2 also just meets
the $r$-I classification with [Fe/H]$_{\mathrm {LTE}}$=$-$1.70, [Eu/Fe]=$+$0.35 and
[Ba/Eu]=$-$0.48~dex.  
Indeed the [Ba/Eu] abundance ratio in all three stars is consistent
with the scaled-solar $r$-process value. 
The lower panel of Fig.\ \ref{fig:eu} shows the abundance ratio
[Ba/Eu] as a function of [Fe/H]$_{\mathrm {LTE}}$.
The average [Ba/Eu] for all the seven stars with both Ba and Eu
measurements is $-$0.54, close to the scaled-solar $r$-process value. 
We note that for these stars, the standard
deviations of the [Ba/Fe] and [Eu/Fe] values are 0.28 and 0.34~dex,
respectively.  However, the standard deviation of the [Ba/Eu] ratio
for these stars is substantially less at 0.13~dex. This may indicate
that while the total production of Ba and Eu is variable, the
nucleosynthetic site(s) involved produce Ba and Eu in very similar
relative amounts.

%
\begin{figure}
   \centering
\includegraphics[width=0.47\textwidth,angle=0.]{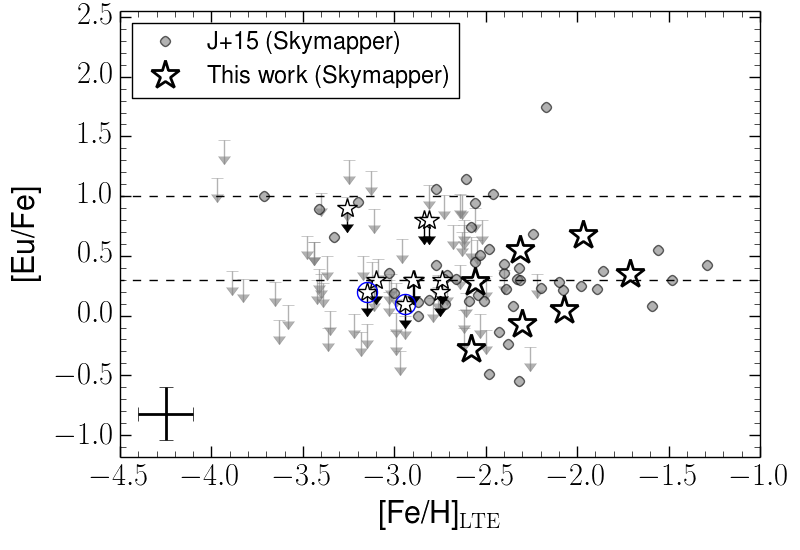}
\includegraphics[width=0.47\textwidth,angle=0.]{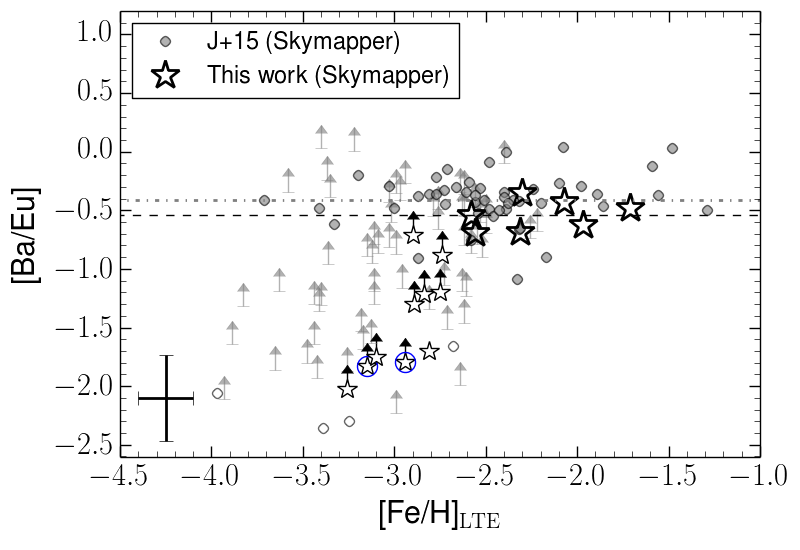}
      \caption{{\it Upper panel}: [Eu/Fe] as a function of
        [Fe/H]$_{\mathrm {LTE}}$ for our sample compared with the 
        \citet{HJ15} sample. In both cases upper limits on [Eu/Fe]
        are indicated by downward pointing arrows. The dashed lines
        are [Eu/Fe]=$+$0.3 and [Eu/Fe]=$+$1 show the
        classification values for $r$-I and $r$-II stars, respectively.   
        The two stars indicated with blue open corcles are the
          two stars with the lowest neutron-capture element abundances.
        {\it Lower panel}: [Ba/Eu] as a function of [Fe/H]$_{\mathrm
          {LTE}}$ for the stars with measured Eu abundances or upper
        limits. Symbols are as in the upper panel. The four open
        circles in the \citet{HJ15} sample are the stars with upper
        limits in both Ba and Eu. One star in our sample also has only
        upper limits, and has been represented with a small star-like
        symbol, without upward limit. 
        The black dashed line and the grey dotted-dashed line are the mean
        [Ba/Eu] abundances in our sample and \citet{HJ15},
        respectively. The size of the {\it y} axis has been kept the
        same as in the upper panel.} 
        \label{fig:eu}
   \end{figure}
%

\section{Summary} \label{Sect5}

We have presented here the results of an analysis of high-resolution
spectra, obtained with the Keck telescope and the HIRES spectrograph,
of 17 candidate extremely metal-poor stars selected from SkyMapper
commissioning-era photometry.  Fourteen of the stars had not
previously been observed at high-dispersion.  We find that, as in
\citet{HJ15}, the candidate selection process, i.e., photometry plus
low-resolution spectroscopy, is robust with almost half of the sample
having \mbox{[Fe/H]$_{\mathrm {LTE}} \leq -$2.8} and with only one `false
positive' -- an EMP-candidate for which [Fe/H] turned out to
exceed [Fe/H]$_{\mathrm {LTE}}$=$-$2.0~dex.  In general, the distribution of
element abundances and abundance ratios for this sample closely mimics
the earlier results of \citet{HJ15} that was based on Magellan/MIKE
high-dispersion spectroscopy of a large sample of SkyMapper
commissioning-era EMP candidates.  Specifically, we find that none of
the present sample can be classified as CEMP stars.  Further, we
confirm the results of \citet{HJ15} that the star
SMSS~J034249.52--284215.8 is an example of the relatively rare class
of objects known as ``Fe-enhanced'' stars -- stars with generally
sub-solar abundance ratios, including for the $\alpha$-elements.  The
star may have originated in a dwarf spheriodal galaxy.  Two further
stars, SMSS~J202659.41--031149.2 and SMSS~J222349.48--114751.1 are
found to be strongly depleted in $n$-capture elements: [Ba/Fe]$<
-$1.6, [Sr/Ba]$\approx -$0.3~dex, and [Eu/Fe]$\lesssim +$0.10-0.15 
joining the similar star identified in \citet{HJ15}, while the star
SMSS~J202400.03--024445.9 is a probable $r$-I star with
[Eu/Fe]=$+$0.68, [Ba/Eu]=$-$0.63 and [Fe/H]$_{\mathrm {LTE}}$=$-$1.95~dex.     

\section*{Acknowledgements}

We thank the anonymous referee for his/her suggestions that improved
the manuscript.
We thank Dr Vini Placco for providing the evolutionary mixing
corrections to the observed carbon abundances. 
SkyMapper research on EMP stars has been supported in part though the
Australian Research Council (ARC) 
Discovery Grant programs DP120101237 and DP150103294 (Lead-CI Da~Costa). 
A.~F.~M., A.~R.~C., and A.~D.~M. have been in part supported by ARC
through the Discovery Early Career Researcher Award 
DE160100851, the Discovery Project DP160100637, and the Future
Fellowship FT160100206, respectively.
M.~A. gratefully acknowledges generous funding from an ARC Laureate
Fellowship (grant FL110100012). Parts of this research were conducted
under the auspices of the Australian Research Council Centre of
Excellence for All Sky Astrophysics in 3 Dimensions (ASTRO 3D) which
is supported through project number CE170100013. 
This project has received funding from the European Union's Horizon
2020 research and innovation programme under the Marie
Sk{\l}odowska-Curie Grant Agreement No. [797100; Beneficiary: AFM]. 

The national facility capability for SkyMapper has been funded through
ARC LIEF grant LE130100104 from the Australian Research Council,
awarded to the University of Sydney, the Australian National
University, Swinburne University of Technology, the University of
Queensland, the University of Western Australia, the University of
Melbourne, Curtin University of Technology, Monash University and the
Australian Astronomical Observatory. SkyMapper is owned and operated
by The Australian National University's Research School of Astronomy
and Astrophysics. The survey data were processed and provided by the
SkyMapper Team at ANU\@. The SkyMapper node of the All-Sky Virtual
Observatory (ASVO) is hosted at the National Computational
Infrastructure (NCI).  Development and support the SkyMapper node of
the ASVO has been funded in part by Astronomy Australia Limited (AAL)
and the Australian Government through the Commonwealth's Education
Investment Fund (EIF) and National Collaborative Research
Infrastructure Strategy (NCRIS), particularly the National eResearch
Collaboration Tools and Resources (NeCTAR) and the Australian National
Data Service Projects (ANDS).

The high dispersion spectra presented herein were obtained at the
W.M. Keck Observatory, which is operated as a scientific partnership
among the California Institute of Technology, the University of
California and the National Aeronautics and Space Administration. The
Observatory was made possible by the generous financial support of the
W.M. Keck Foundation. 

The authors wish to recognize and acknowledge the very significant
cultural role and reverence that the summit of Mauna Kea has always
had within the indigenous Hawaiian community.  We are most fortunate
to have the opportunity to conduct observations from this mountain. 

We also acknowledge the traditional owners of the land on which the
SkyMapper telescope stands, the Gamilaraay people, and pay our
respects to elders past and present.

\end{document}